\documentclass[12pt]{article}
\usepackage{graphicx, psfrag, epsf}
\usepackage{caption, subcaption, float, color, multirow, bbm}
\usepackage[colorlinks=true, urlcolor=blue,linkcolor=blue, citecolor=magenta, breaklinks=true]{hyperref}
\usepackage{natbib}
\usepackage{empheq}
\usepackage{enumerate}
\usepackage{dsfont}
\usepackage{amssymb}
\usepackage{amsthm}
\usepackage{mathrsfs}
\usepackage{bigints}
\usepackage{fancyhdr}
\usepackage{empheq}
\usepackage{mathtools}
\usepackage{bm}
\usepackage{stmaryrd}
\usepackage{amsmath}
\usepackage{amsfonts}
\usepackage{authblk}
\usepackage{sectsty}
\usepackage{times}
\usepackage[plain,noend]{algorithm2e}
\newtheorem{thm}{Theorem}

\newtheorem{cor}{Corollary}

\newtheorem{rk}{Remark}
\newtheorem{theorem}[thm]{Theorem}
\newtheorem{assumption}[]{Assumption}

\numberwithin{equation}{section}

\newcommand{\bs}{\boldsymbol}

\newcommand{\nc}{\newcommand}
\nc{\dps}{\displaystyle}
\nc{\tr}{\text{tr}}

\DeclareMathOperator*{\argmin}{argmin}

\allowdisplaybreaks
\def\boxit#1{\vbox{\hrule\hbox{\vrule\kern6pt
\vbox{\kern6pt#1\kern6pt}\kern6pt\vrule}\hrule}}

\begin{document}

\date{}
\title{\textbf{Data-Driven Uniform Inference for General
Continuous Treatment Models via Minimum-Variance Weighting}\footnote{The authors are alphabetically ordered.}}
	\author[$^\dagger$]{Chunrong Ai \thanks{E-mail: 
		\texttt{chunrongai@cuhk.edu.cn} }}
\author[$^\ddagger$]{Wei Huang \thanks{E-mail: 
		\texttt{wei.huang@unimelb.edu.au} }}
\author[$^\star$] {Zheng Zhang \thanks{%
		E-mail: \texttt{zhengzhang@ruc.edu.cn}}}

\affil[$\dagger$]{School of Management and Economics, Chinese University of Hong Kong, Shenzhen}	
\affil[$\ddagger$]{	School of Mathematics and Statistics, University of Melbourne}	
\affil[$\star$]{Center for Applied Statistics, Institute of Statistics \& Big Data, Renmin University of China}

\global\long\def\e{\mathbb{E}}%

\maketitle

\begin{abstract}
\cite{ai2021unified} studied the estimation of a general dose-response function (GDRF) of a continuous treatment that includes the average dose-response function, the quantile dose-response function, and other expectiles of the 
dose-response distribution. They specified the GDRF as a \emph{parametric} function of the treatment status only and  proposed a weighted regression with the weighting function estimated using the maximum entropy approach. This paper specifies the GDRF as a nonparametric function of the treatment status,  proposes a weighted local linear regression for estimating GDRF, and develops a bootstrap procedure for constructing the \emph{uniform} confidence bands. We propose stable weights with minimum sample variance while eliminating the sample association between the treatment and the confounding variables. The proposed weights admit a closed-form expression, allowing them to be computed efficiently in the bootstrap sampling. Under certain conditions, we derive the uniform Bahadur representation for the proposed estimator of GDRF and establish the validity of the corresponding uniform confidence bands.  A fully data-driven approach to choosing the undersmooth tuning parameters and a data-driven bias-control confidence band are included. A simulation study and an application demonstrate the usefulness of the proposed approach.  
\end{abstract}

\sloppy

\renewcommand{\baselinestretch}		{1}

{\normalsize 
%%%%%%%%%%%%%%%%%%%%%%%%%%%%%%%%%%%%%%%%%%%%%%%%%%%%%%%%%%%%%%%%%%%%%%%%%%%%%%
}

{\normalsize \if00 {\ }}

{\normalsize \ }

{\normalsize \ }

{\normalsize \ \fi
}

{\normalsize \if10 {\ \bigskip \bigskip \bigskip }}

\begin{center}
{\normalsize \ {\LARGE \textbf{Non-parametric Uniform Inference for General
Treatment Models}} }
\end{center}

{\normalsize \medskip \fi
}


{\normalsize \renewcommand\thefootnote{\arabic{footnote}} }

\section{Introduction}
 \cite{ai2021unified} studied the estimation of a general dose-response function (GDRF) that
encompasses the average dose-response function (ADRF), the quantile
dose-response function (QDRF), and other expectiles (percentiles) of the dose-response distribution. They specified the GDRF as a \emph{parametric} function of the treatment status
only. They proposed a weighted regression estimator with the weight function estimated
by the maximum entropy approach. Under some sufficient conditions, they
established large sample properties of the estimated model parameters, e.g., root $n$ asymptotic normality and  efficiency bounds. This paper specifies the GDRF as a nonparametric function of the treatment
status and proposes a weighted local linear regression.
Under some sufficient conditions, we establish the asymptotic properties of
the nonparametric estimator of the GDRF and
the \emph{uniform} confidence bands. These results are helpful because a false
parameterization of the GDRF could lead to an
erroneous conclusion, and the uniform confidence bands enable
a more powerful inference of the treatment-effect parameter. For example, it is of interest to test whether the mother's age, varying within a certain age range (a continuous treatment), has a significant effect on the baby’s birth weight, or to analyze the effect of the amount of commercial advertisements (a continuous treatment) on sales.

Our paper also contributes to the literature of uniform nonparametric inference. Different from existing studies of uniform nonparametric inference for the probability density function \citep{chernozhukov2014anti,kato2018uniform}), the cumulative distribution function \citep{adusumilli2020inference}, and the conditional mean regression function \citep{horowitz2017nonparametric,kato2019uniform,li2020uniform}, where the parameters of interest have closed-form expressions and do not depend on additional nuisance parameters, our GDRF is defined as an \emph{implicit} solution of a general weighted optimization problem, where the weighting function is unknown. The criteria function could be possibly non-smooth (e.g., the quantile loss function). These significant  differences give rise to new challenges, making previous methods inapplicable.  To overcome the problems above, we propose a novel method for estimating weights by minimizing their sample variance, subject to an increasing number of moment restrictions.  The minimum-variance weights admit a closed-form expression that can be computed stably and efficiently. Then, we develop a uniform inference procedure via a weighted bootstrap on the criteria function for the estimator, and a fully data-driven inference procedure is also included. We derive a uniform Bahadur representation for our proposed estimator of GDRF and validate the construction of uniform confidence bands.  Monte Carlo simulation studies and a real data application demonstrate the usefulness of the proposed method.

The paper is organized as follows. Section \ref{sec:framework} introduces the GDRF parameters and the goal of the paper. Section~\ref{sec:ourinferenceframework} presents our methodology, including the nonparametric estimation and inference for GDRF parameters. Section \ref{sec:asymptotics} establishes the theoretical properties of the estimated GHTE and the theoretical validity of our uniform inferences. Section \ref{sec:NumericalDetails} gives practical suggestions including a data-driven undersmooth tuning parameters selector and data-driven bias-control confidence bands. Section \ref{sec:numerical} reports a simulation study, while Section \ref{sec:empirics} presents an empirical application. All technical proofs and additional simulation results are included in the supplemental material.

\section{Basic Framework} \label{sec:framework} 
Let $T$ denote the observed continuous treatment status variable, with
support $\mathcal{T}\subset \mathbb{R}$ and a probability density function $f_{T}(t)$. Let $Y^{\ast }(t)$ denote the potential response when treatment $t$ is assigned. The probability density of $Y^{\ast }(t)$ exists, denoted by $f_{Y^{\ast }(t)}$, and is continuously
differentiable. Let $Y=Y^{\ast }(T)$ denote the observed response. Let $\mathcal{L}(\cdot )$ denote a nonnegative, strictly convex, and possibly non-smooth loss function satisfying $\mathcal{L}(0)=0$ and $\mathcal{L}(v)\geq 0$ for all $v\in \mathbb{R}$. The derivative of $\mathcal{L}(\cdot )$, denoted by $\mathcal{L}^{\prime }(\cdot )$, exists almost everywhere and is non-constant. Let $\boldsymbol{X}\in \mathbb{R}^{d_{X}}$, for some integer $d_{X}\geq 1$, denote a vector of observable covariates with support $\mathcal{X}$. 

We consider the \emph{general dose-response function} (GDRF), $g(\cdot)$, defined as a unique solution to the following convex optimization problem: 
\begin{equation}\label{model:ini}
g(t):=\argmin_{a\in\mathbb{R}}\mathbb{E}%
\left[ \mathcal{L}\left\{ Y^{\ast }(t)-a\right\}\right],
\end{equation}
for every $t\in\mathcal{T}$. The \emph{general continuous treatment-effect} (GCTE), is defined as $\tau(t_{1},t_{0}):=g(t_{1})-g(t_{0})$ when the treatment status is changed discretely from $T=t_{0}$ to $T=t_{1}$ or the \emph{general marginal treatment effect} (GMTE) $g'(t)$, the derivative of $g(t)$, when the treatment status is changed marginally.  To identify $g(\cdot)$ in \eqref{model:ini} from the observables, throughout the paper, we impose the following widely accepted unconfoundedness condition \citep{hirano2004propensity}.
\begin{assumption}[\emph{Unconfoundedness}]\label{as:TYindep} 
	For all $t\in \mathcal{T}$, given $\boldsymbol{X}$, $T$ is independent of $Y^{\ast }(t)$, i.e., $Y^{\ast}(t)\perp T|\boldsymbol{X}$, for all $t\in \mathcal{T}$. 
\end{assumption}

Let $f_{T|X}$ denote the conditional probability density function of $T$ given the
observed covariates $\boldsymbol{X}$, which is called the generalized propensity score (GPS) in \cite{hirano2004propensity}. Under Assumption~\ref{as:TYindep}, the GDRF, $g(t),$ solves 
\begin{equation}\label{eq:g0_min}
g(t)=\argmin_{a\in\mathbb{R}}\mathbb{E}\left[ \pi(T,\boldsymbol{X})\mathcal{L}\left\{ Y-a\right\}|T=t\right].
\end{equation}
where 
\begin{equation}\label{eq:pidef}
\pi(T,\boldsymbol{X}):=\frac{f_{T}(T)}{f_{T|X}(T|\boldsymbol{X})}
\end{equation}
is the \emph{stabilized weight function} \citep{robins2000marginal,ai2021unified}. The proof of \eqref{eq:g0_min} is presented in Appendix \ref{app:iden_weighted}.

Model \eqref{model:ini} encompasses a variety of continuous treatment effect parameters of interest. For example, with $\mathcal{L}(v)=v^{2}$ \eqref{model:ini} gives $g(t)=\mathbb{E}\{Y^{\ast }(t)\}$the average dose-response function (ADRF), which is widely studied in the existing literature, see \cite{kennedy2017non,colangelo2020double,huang2022nonparametric}. With $\mathcal{L}(v)=v\{q -\mathds{1}(v\leq 0)\}$ for some $q \in (0,1)$, model \eqref{model:ini} gives the quantile dose-response function (QDRF) $g(t)=F_{Y^{\ast }(t)}^{-1}(q)=\inf \{\rho:\mathbb{P}\{Y^{\ast
}(t)\geq \rho\}\leq q \}.$  It is worth noting that all papers above only focused on the pointwise nonparametric inference of ADRF, we are unaware of any literature studying its uniform asymptotic properties, let alone more general results on the uniform inference for GDRF. This paper bridges the gap by developing nonparametric estimators for the general causal effect functions $g(\cdot )$, $\tau(\cdot,\cdot)$ and $g'(\cdot )$, establishing their uniform asymptotic results, and constructing uniform confidence bands. The uniform confidence bands are instrumental as we can use them to test the null: 
\begin{align}
 &H_{0}:g (t)=g_0(t)\ \text{for all $t\in \mathcal{T}$;} \label{H0:g}\\
& H_{0}:\tau (t_1,t_0)={\tau}_0(t_1,t_0)\ \text{for all $(t_0,t_1)\in \mathcal{T}\times\mathcal{T}$;} \label{H0:tau}
\end{align}
where ${g}_0(t)$ and ${\tau}_0(t_1,t_0)$ are assumed to be known functions, as well as the following null
\begin{align}
H_{0}:g' (t)=0\ \text{for all $t\in \mathcal{T}$.} \label{H0:g'}
\end{align}

We first brief our idea of constructing the nonparametric uniform band for $g$. The bands for $g'$ and $\tau$ can be established in the same way. Based on \eqref{eq:g0_min}, once we have an estimator $\widehat{\pi}$ for $\pi$,  we consider estimating $g(t)$ and $g'(t)$ simultaneously using a local linear regression: \footnote{Generally, $\widehat{g}$ and $\widehat{\tau}$ lack explicit expressions, requiring iterative numerical methods for estimation. For instance, to calculate quantile treatment effects, the iteratively reweighted least squares algorithm by \cite{lejeune1988quantile} is applicable.}
 \begin{align}\label{def:ghat}
	(\widehat{g}_h(t),\widehat{g'}_h(t)) 
	=\argmin_{(\theta_1,\theta_2)\in\mathbb{R}^2}\sum_{i=1}^{N}\widehat{\pi }(T_{i},\boldsymbol{X}_{i})\mathcal{L}%
	\left\{ Y_{i}-\theta_1-\theta_2(T_i-t) \right\}\mathcal{K}_{h}\left( T_{i}-t\right), 
\end{align}
where $\mathcal{K}_{h}\left( T_{i}-t\right):=\mathcal{K}\left((T_{i}-t)/h\right)$, $\mathcal{K}(\cdot)$ is a kernel function, and $h$ is a bandwidth.  Consequently, we can estimate the GCTE, $\tau(t_1,t_0)$ by $\widehat{\tau}(t_1,t_0):= \widehat{g}_h(t_1) - \widehat{g}_h(t_0)$. Let $$
Z_{g,N}(t) := \frac{\widehat{g}_h(t) - g(t)}{\sigma_{g,N}(t,h)}\,,
$$
and
$$
I_{g,N}(t) := \left[\widehat{g}_h(t) - c_{N}(\alpha)\cdot \sigma_{g,N}(t,h),\widehat{g}_h(t)+ c_{N}(\alpha)\cdot \sigma_{g,N}(t,h)\right],
$$
where $\sigma_{g,N}(t,h)$ is the standard deviation of the limiting distribution of $\widehat{g}_h(t)$ , and $c_{N}(\alpha)$ is the $(1-\alpha)$-quantile of the limiting distribution of $\sup_{t\in\mathcal{T}}|Z_{g,N}(t)|$. Then $\{I_{g,N}(t), t\in \mathcal{T}\}$ is a $(1-\alpha)$ nonparametric uniform band for $g$, in the sense that 
\begin{equation*}
	\lim_{N\rightarrow\infty}\mathbb{P}\{g(t)\in I_{g,N}(t), \forall t\in\mathcal{T}\} = 1-\alpha\,.
\end{equation*}

To determine the critical value $c_N(\alpha)$, we consider estimating $c_N(\alpha)$ by a weighted bootstrap algorithm \citep{zhang2020quantile}. Specifically, let $B$ be the times of bootstrap, the $b^{th}$ bootstrap version of $(\widehat{g}_h,\widehat{g'}_h)$ is 
\begin{align}
	(\widehat{g}^{(b)}_h(t),\widehat{g'}^{(b)}_h(t))
	=\argmin_{(\theta_1 ,\theta_2)}\sum_{i=1}^{N}\xi _{i}^{(b)}{\widehat{\pi }^{(b)}}(T_{i},\boldsymbol{X}%
	_{i})\mathcal{L}\left\{ Y_{i}-\theta_1 -(T_{i}-t)\theta_2 \right\} \mathcal{%
	K}_{h}\left( T_{i}-t\right),\label{eq:gtau_bootstrap}
\end{align}
where $\xi_i^{(b)}$'s are independent \emph{positive} random variables with mean one and variance one, and $b=1,...,B$. For example, we can take a set of i.i.d. $2\times\text{Bernoulli}(1/2)$ random variables. The weights, $\widehat{\pi }^{(b)}$'s, are the $b^{th}$ bootstrap version of $\widehat{\pi}$. We then estimate $c_N(\alpha)$ by the $(1-\alpha)$ empirical quantile of $\{\widehat{Z}^{(b)}_{g,N}\}_{b=1}^B$, where
$$
\widehat{Z}^{(b)}_{g,N}:=\frac{\widehat{g}^{(b)}_h(t)-\widehat{g}_h(t)}{\widehat{\sigma}_{g,N}(t,h)},
$$
where $\widehat{\sigma}_{g,N}(t,h)$ is an estimator of $\sigma_{g,N}(t,h)$.

To establish the validity of our uniform confidence band both practically and theoretically, we need to complete the following three parts:
\begin{enumerate}
	\item [Part I:] Propose an estimator, $\widehat{\pi}$, and its bootstrap versions, $\{\widehat{\pi }^{(b)}\}_{b=1}^B$, for $\pi$, that can be easily and efficiently computed.  As highlighted in \eqref{eq:gtau_bootstrap}, constructing uniform confidence bands necessitates calculating bootstrap weights $\{\widehat{\pi }^{(b)}\}_{b=1}^B$ for $B$ times, which can be computationally intensive.\footnote{\cite{ai2021unified} proposed an estimator for $\pi$ by the maximum entropy approach. However, their estimator does not have a closed-form expression and needs to be computed using numerical methods, which is not suitable for bootstrap repetition in practice.} To enhance computational efficiency and feasibility, in Section \ref{sec:ourinferenceframework}, we propose a new estimator for $\pi$ and its bootstrap counterpart that provide closed-form solutions, significantly simplifying the inference process.
	\item [Part II:] Establish the uniform  Bahadur representations for $(\widehat{g}_h,\widehat{g'}_h)$ and $(\widehat{g}^{(b)}_h,\widehat{g'}^{(b)}_h)$, for $b=1,\ldots,B$. With the uniform  Bahadur representations, by applying Gaussian approximation theory of suprema of empirical processes  (\cite{chernozhukov2014gaussian}) and the anti-concentration property of tight Gaussian processes (\cite{chernozhukov2014anti}),  we shall establish the validity of our bootstrap uniform confidence bands in Section \ref{sec:asymptotics}.
    \item [Part III:] Provide practical details, including the data-driven undersmooth tuning parameters and the construction of data-driven bias-control confidence bands with theoretical validity in Section~\ref{sec:NumericalDetails}.
\end{enumerate}

\begin{rk}
\cite{ai2021unified} studied a parametric-response function, $%
g(t,\boldsymbol{\beta }_{0}),$ with the finite-dimensional parameter $%
\boldsymbol{\beta }_{0}\in \mathbb{R}^{p}$ solving the following
optimization problem: 
\begin{equation*}
\boldsymbol{\beta }_{0}=\argmin_{\boldsymbol{\beta }}\mathbb{E}\left[ {\pi }(T,\boldsymbol{X})\mathcal{L}\{Y-g(T;\boldsymbol{\beta })\}\right] ,
\end{equation*}%
where $g(\cdot ,\cdot )$ is a known function. Such a parametric-response function could be misspecified, leading to a false conclusion. 
\end{rk}

\begin{rk}
An alternative approach to the local linear regression \eqref{def:ghat} of estimating GDRF is the sieve generalized empirical likelihood (GEL)
proposed in \cite{chen2019penalized}. Note that $g(T)$ solves the following
conditional-moment restriction, 
\begin{equation*}
\mathbb{E}\left[ \pi(T,\boldsymbol{X})\mathcal{L}^{\prime
}(Y-g(T))|T\right] =0,
\end{equation*}%
which implies 
\begin{equation*}
\mathbb{E}\left[ \pi(T,\boldsymbol{X})\mathcal{L}^{\prime
}(Y-g(T))v_{K_{1}}(T)\right] =0,
\end{equation*}%
where $v_{K_{1}}(T)$ is a $K_{1}$-dimensional approximating sieve. Let
$s(\cdot )$ be a strictly concave and twice-continuously differentiable
function, with Lipschitz-continuous second derivative and satisfy $s^{\prime
}(0)=s^{\prime \prime }(0)=-1$. Following \cite{chen2019penalized}, the
sequential GEL estimator of $g(t)$ is given by 
\begin{equation*}
\widehat{g}_{GEL}(t)=\widehat{\alpha }_{K_{0}}^{\top }v_{K_{0}}(t),
\end{equation*}%
where $\widehat{\alpha }_{K_{0}}\in \mathbb{R}^{K_{1}}$ solves the following
saddle point optimization problem: 
\begin{equation*}
\widehat{\alpha }_{K_{0}}=\arg \min_{\alpha \in \mathbb{R}%
^{K_{0}}}\max_{\gamma \in \mathbb{R}^{K_{1}}}\frac{1}{N}\sum_{i=1}^{N}s\left%
\{ \widehat{\pi }(T_{i},\boldsymbol{X}_{i})\mathcal{L}^{\prime
}(Y_{i}-\alpha ^{\top }v_{K_{0}}(T_{i}))\cdot \gamma ^{\top
}v_{K_{1}}(T_{i})\right\} -s(0).
\end{equation*}%
However, with a general $\mathcal{L}^{\prime
}(\cdot)$ that could possibly be non-differentiable and an estimated $\widehat{\pi }$, the uniform inference for $g(\cdot)$ and $g'(\cdot)$ is more difficult to develop. We will pursue this
extension in a future project.
\end{rk}

\begin{rk} \label{rk:DR}
	Doubly-robust treatment effect estimation has gained popularity. For continuous treatment, \cite{kennedy2017non} propose a nonparametric estimator for the ADRF  based on a doubly-robust representation, i.e.  $\mathbb{E}[Y^*(t)]=\mathbb{E}[\xi(T,\boldsymbol{X},Y)|T=t]$, where
	$$\xi(T,\boldsymbol{X},Y):= \pi(T,\bs{X})\{Y - m(T,\boldsymbol{X})\}+\int m(T,\boldsymbol{x})f_X(\boldsymbol{x})d\boldsymbol{x}$$
	and $m(T,\boldsymbol{X}):=\mathbb{E}[Y|T,\boldsymbol{X}]$ is the outcome regression function.  They then estimate $\mathbb{E}[Y^*(t)]$ by regressing the estimated pseudo-outcome $\xi(T,\boldsymbol{X},Y)$  onto $T=t$ through the local linear polynomial, i.e. $\widehat{\theta}_1(t)$, where
	\begin{align*}
		&(\widehat{\theta}_1(t),\widehat{\theta}_2(t)):=\arg\min_{(\theta_1,\theta_2)\in\mathbb{R}^2}\sum_{i=1}^N \mathcal{K}_h\left(T_i-t\right)\left\{\widehat{\xi}(T_i,\boldsymbol{X}_i,Y_i)-\theta_1-(T_i-t)\theta_2\right\}^2,\\
		&\widehat{\xi}(T,\boldsymbol{X},Y) := \widehat{\pi}(T,\bs{X})\{Y - \widehat{m}(T,\boldsymbol{X})\}+ \frac{1}{N}\sum_{i=1}^N\widehat{m}(T,\boldsymbol{X}_i),
	\end{align*}
	where $\widehat{\pi}(\cdot)$ and $\widehat{m}(\cdot)$ are some consistent nonparametric estimators for $\pi(\cdot)$ and $m(\cdot)$. \citet[Theorem 2]{kennedy2017non} establishes the pointwise asymptotic result for $\widehat{\theta}_1(t)$ for every fixed $t\in\mathcal{T}$ and its double robustness. However, their method relies on the closed form of a doubly-robust representation of ADRF, which makes it challenging to extend it to estimate our GDRF $g(\cdot)$ , which is an implicitly defined causal function. Moreover, the investigation of the uniform inference for $\mathbb{E}[Y^*(t)]$ over $t\in\mathcal{T}$ remains unknown. We will pursue  the doubly-robust estimation of $g(\cdot)$ and its uniform inference in future work.    
\end{rk}

\section{Minimum-Variance Weights and Uniform Inference Algorithm}\label{sec:ourinferenceframework}

Several methods may be used to estimate $\pi$. For example, one may estimate the numerator and denominator in \eqref{eq:pidef} separately and then form a ratio estimator. However, such an estimator is known to be unstable due to its sensitivity to the denominator's estimation \citep{kang2007demystifying}. To address the instability issue, \cite{ai2021unified} proposed a direct estimation of $\pi$ by minimizing the Kullback-Leibler (KL) divergence between $\{N^{-1}\pi(T_i,\boldsymbol{X}_i)\}_{i=1}^N$ and the uniform design weights $\{N^{-1}\}_{i=1}^N$ subject to an expanding set of product-moment constraints. This method, while addressing the instability issue, does not offer a closed-form solution and requires iterative numerical methods for computation.  To overcome the problems above, we extend the method of \cite{ai2021unified} by replacing the KL-divergence with the squared loss such that the estimated weights admit a closed-form expression that can be stably and efficiently computed.

We notice that, for any suitable function $u(t,\boldsymbol{x})$, the weight
function $\pi(t,\boldsymbol{x})$ satisfies the moment restriction, 
\begin{equation}
\mathbb{E}\left\{ \pi(T,\boldsymbol{X})u(T,\boldsymbol{X})\right\}
=\int u(t,\boldsymbol{x})f_{T}(t)f_{X}(\boldsymbol{x})dtd\boldsymbol{x}.
\label{moment1}
\end{equation}%
Equation (\ref{moment1}) identifies $\pi(t,\boldsymbol{x})$ so we can
estimate $\pi(T_{i},\boldsymbol{X}_{i})$ by solving the sample analogue
of (\ref{moment1}). The challenge is that ( \ref{moment1}) has an infinite
number of restrictions. It is impossible to impose an infinite number of
restrictions on a finite number of sample observations. To overcome this
difficulty, we approximate a functional space by a sequence of
finite-dimensional sieve spaces. Specifically, let $u_{K}(T,\boldsymbol{X}%
)=(u_{K,1}(T,\boldsymbol{X}),\ldots ,u_{K,K}(T,\boldsymbol{X}))^{\top }$
denote the \emph{approximation sieves}, such as B-splines and power series 
\citep[see][for more discussion on sieve
	approximation]{chen2007large}. $\pi(T,\boldsymbol{X})$ also
satisfies 
\begin{equation}
\mathbb{E}\left\{ \pi(T,\boldsymbol{X})u_{K}(T,\boldsymbol{X})\right\}
=\int u_{K}(t,\boldsymbol{x})f_{T}(t)f_{X}(\boldsymbol{x})dtd\boldsymbol{x}%
\,.  \label{sievemoment}
\end{equation}%

Noting that $\mathbb{E}\{\pi(T,\boldsymbol{X})\}=1$ and $\text{var}\{\pi(T,\boldsymbol{X})\} = \mathbb{E}[\{\pi(T,\boldsymbol{X})-1\}^2]$, with the moment restrictions \eqref{sievemoment},  we propose the nonparametric \emph{minimum-variance weights} $\widehat{\pi}_{K}(T_i,\boldsymbol{X}_i)$: 
\begin{equation}
\left\{ 
\begin{array}{cc}
& \left\{ \widehat{\pi}_{K}(T_i,\boldsymbol{X}_i)\right\} _{i=1}^{N}=\argmin_{\pi_i}\sum_{i=1}^{N}(\pi
_{i}-1)^2/2 \\[2mm] 
& \text{subject to}\ \frac{1}{N}\sum_{i=1}^{N}\pi _{i}u_{K}(T_{i},%
\boldsymbol{X}_{i})=\frac{1}{N(N-1)}\sum_{j=1,j\neq
i}^{N}\sum_{i=1}^{N}u_{K}(T_{i},\boldsymbol{X}_{j}).
\end{array}
\right.  \label{E:cm2}
\end{equation}
Note that when $u_{K,1}(T,\boldsymbol{X})=1$ is included in the sieve basis, \eqref{E:cm2} guarantees that $\sum^N_{i=1}\widehat{\pi}_K(T_i,\boldsymbol{X}_i)/N = 1$. In Appendix \ref{app:pihat}, we show that the primal problem \eqref{E:cm2} has a closed-form solution, 
\begin{align}
	\widehat{\pi }_{K}(t,\boldsymbol{x})=&- \left( \frac{1}{N}\sum_{i=1}^{N}u_{K}\left( T_{i},\boldsymbol{X}_{i}\right) -  \frac{1}{N(N-1)}\sum_{i=1}^{N}\sum_{j=1,i\neq j}^{N}u_{K}\left( T_{i},\boldsymbol{X}_{j}\right)\right) ^{\top } \notag \\
	& \times \left( \frac{1}{N}\sum_{i=1}^{N}u_{K}\left( T_{i},\boldsymbol{X}_{i}\right) u_{K}^{\top }\left( T_{i},\boldsymbol{X}_{i}\right) \right)^{-1}u_{K}\left( t,\boldsymbol{x}\right) +1. \label{def:pihat}
\end{align}
Correspondingly, the bootstrap weights are
\begin{align}
	\widehat{\pi }^{(b)}_{K}(t,\boldsymbol{x})=& -\left( \frac{1}{N}\sum_{i=1}^{N}\xi^{(b)}_i u_{K}\left( T_{i},\boldsymbol{X}_{i}\right) -  \frac{1}{N(N-1)}\sum_{i=1}^{N}\sum_{j=1,i\neq j}^{N}u_{K}\left( T_{i},\boldsymbol{X}_{j}\right)\right) ^{\top } \notag \\
	& \times \left( \frac{1}{N}\sum_{i=1}^{N} \xi^{(b)}_i u_{K}\left( T_{i},\boldsymbol{X}_{i}\right) u_{K}^{\top }\left( T_{i},\boldsymbol{X}_{i}\right) \right)^{-1}u_{K}\left( t,\boldsymbol{x}\right) +1. \label{def:pihatb}
\end{align}

We now introduce the procedure of constructing the uniform confidence bands for $g$. For $g'$, and $\tau$, we can build the uniform confidence bands using the same algorithm:

\begin{enumerate}
\item  Compute the estimator, $\widehat{g}_h(t)$ in \eqref{eq:g0_min} , with $\widehat{\pi}$ replaced by $\widehat{\pi}_K$ in \eqref{def:pihat}, for a suitably fine grid over its support, $\mathcal{T}$. 

\item  Consider a sufficiently large integer, $B$. Compute the bootstrap estimators, $\widehat{g}^{(b)}_h(t)$ in \eqref{eq:gtau_bootstrap} with $\widehat{\pi}^{(b)}$ replaced by $\widehat{\pi}_K^{(b)}$ in \eqref{def:pihatb}, over the same grid, for $b=1,...,B$, 

\item  Calculate a consistent estimate of the asymptotic variance, $\widehat{\sigma }_{g,N}(t)$. For example, we can estimate $\sigma _{g,N}(t)$ by the bootstrap standard deviation, i.e., the standard deviation of $\{\widehat{g}^{(b)}_h(t)-\widehat{g}_h(t)\}_{b=1}^{B}$. Alternatively, we can take the bootstrap normalized inter-quartile range, i.e., $\widehat{\sigma }_{g,N}(t) = \{q_{0.75}(t) - q_{0.25}(t)\}/\{z_{0.75}(t) - z_{0.25}(t)\}$, where $q_p(t)$ and $z_p(t)$ are the $p$th quantile of $\{\widehat{g}^{(b)}_h(t)-\widehat{g}_h(t)\}_{b=1}^{B}$ and that of $N(0,1)$, respectively.
%Estimate $\sigma _{1,N}(t,z)$ by the sample standard deviation of $\sqrt{Nh}h_{0}\{\widehat{\tau }^{(b)}(t|z)-\widehat{\tau }(t|z)\}_{b=1}^{B}$, denoted by $\widehat{\sigma }_{1,N}(t,z)$. Estimate $\sigma _{2,N}(t_{1},t_{0},z)${\normalsize \ by the sample standard deviation of $\sqrt{Nh}\{\widehat{\tau }^{(b)}(t_{1},t_{0}|z)-\widehat{\tau }(t_{1},t_{0}|z)\}_{b=1}^{B}$, denoted by $\widehat{\sigma }_{2,N}(t_{1},t_{0},z)$. }

\item  Compute  $M_{g,b}:=\sup_{t\in \mathcal{T}}|\widehat{Z}^{(b)}_{g,N}(t)|$, for $b=1,\ldots,B$, where the maximum over the chosen grid points approximates the supremum. 

\item  Given a confidence level $1-\alpha $, find the empirical $%
1-\alpha $ quantile of the set of numbers $\{M_{g,b}:b=1,...,B\}$, which is denoted by $\widehat{C}%
_{g,\alpha }$.

\item  The  two-sided uniform confidence bands for $g$ are constructed as
\begin{align*}
	\widehat{I}_{g,N}:=\left\{ \left( \widehat{g}_h(t)-\widehat{C}_{g,\alpha }\widehat{\sigma}_{g,N}(t,h),\widehat{g}_h(t)+\widehat{C}_{g,\alpha }\widehat{\sigma}_{g,N}(t,h)\right) :t\in \mathcal{T}\right\}\,.
\end{align*}
\end{enumerate}
The confidence bands for $g'(\cdot)$, and $\tau(\cdot,\cdot)$, denoted by $I_{g',N}$ and $I_{\tau,N}$, can be built by replacing $\{\widehat{g}_h, \widehat{g}^{(b)}_h\}_{b=1}^B$ in the above algorithm by $\{\widehat{g'}_h, \widehat{g'}^{(b)}_h\}_{b=1}^B$ and $\{\widehat{\tau}(t_1,t_0): = \widehat{g}_h(t_1)-\widehat{g}_h(t_0), \widehat{\tau}^{(b)}: = \widehat{g}^{(b)}_h(t_1)-\widehat{g}^{(b)}_h(t_0)\}_{b=1}^B$, respectively.

\begin{rk} 
	The minimum variance weight in \cite{yiu2018covariate} is parametric. Note that the association between $T$ and $\bs{X}$ as characterized by $f_{T|X}(T|\bs{X})$ can be eliminated after being weighted by $\pi(T,\boldsymbol{X})$, in the sense that $\pi(T,\bs{X})f_{T|X}(T|\bs{X})=f_T(T)$. The authors consider parametric models $T\sim f_T(T;\alpha)$ and $T|\bs{X} \sim f_0(T|\bs{X};\beta=(\beta_b, \beta_d))$where $\beta_d$ refers to the parameters describing the dependence between $T$ and $\bs{X}$ and $\beta_b$ are the parameters for the baseline distribution. They assume $f_0(T|\bs{X};\beta_b, \beta_d=0)=f_T(T;\alpha)$. Then, given a set of weights $\boldsymbol{\pi}=(\pi_1,...,\pi_N)$, $\beta_d$ can be estimated by solving a weighted score equation
	$$
	\sum_{i=1}^{N}\pi_i\frac{\partial}{\partial  \beta(\bs{\pi})}\log f_{0}(T_i|\bs{X}_i;\beta(\bs{\pi}))\bigg|_{\beta_b(\bs{\pi})= \alpha}=0\,.
	$$
	Their idea is that, if the weights $\bs{\pi}$ satisfy the condition of covariate association elimination as $\pi(T,\boldsymbol{X})$ does, then $\beta_d = 0$ should be the solution to the weighted score equation.
	They thus define the minimum-variance weight estimator as follows:
	\begin{align*}
		\left\{ 
		\begin{array}{cc}
			&\{\widehat{\pi}_i\}_{i=1}^N=\arg\min 
			\sum_{i=1}^{N}\left\{\pi_i-1\right\}^2/2\\[2mm] 
			& \text{subject to}\ \sum_{i=1}^{N}\pi_i\frac{\partial}{\partial  \beta(\boldsymbol{\pi})}\log f_{0}(T_i|\beta(\boldsymbol{\pi})^{\top}{\boldsymbol{X}}_i)\bigg|_{\beta_b(\boldsymbol{\pi})=\widehat{\alpha}, \beta_d(\boldsymbol{\pi})=0}=0,\\& \frac{1}{N}\sum_{i=1}^N\pi_i=1, \  \pi_i>0 \ (i=1,...,N),
		\end{array}
		\right.  
	\end{align*}
where $\widehat{\alpha}$ is obtained by fitting the model $f_T(T_i;\alpha)$ to the observed data. However, \cite{yiu2018covariate} do not investigate theoretical properties about the weights  $\{\widehat{\pi}_i\}_{i=1}^N$ as well as the resulted estimator of treatment effect. In contrast, we propose nonparametric minimum-variance weights and develop the theoretical results on both the weights and the resulting treatment effect estimator.  
\end{rk}

\section{Theoretical Properties} \label{sec:asymptotics} 
In this section, we establish the uniform asymptotic properties of our estimators and validate our uniform confidence bands by finishing Part II in Section~\ref{sec:framework}. To aid exposition, we introduce some notation. For any matrix $%
\boldsymbol{A}$, $\Vert \boldsymbol{A}\Vert:=\{\tr(\boldsymbol{A}\boldsymbol{A}^{\top})\}^{1/2}$ denotes the Frobenius norm of $\boldsymbol{A}$. For any measurable function $f(\cdot)$ and an arbitrary probability measure $Q$, let $\Vert f\Vert_{Q,2}=\left( \int |f(\boldsymbol{o})|^{2}dQ(\boldsymbol{o})\right) ^{1/2}$ denote the $L^{2}(Q)$ norm. For any sequences $a_{N},b_{N}\in \mathbb{R}^{+}$, %$a_{N}\lesssim b_{N}$ means $a_{N}\leq Cb_{N}$ for some constant $C>0$ for all $N$,
$a_{N}\prec b_{N}$ means $a_{N}/b_{N}\rightarrow 0$ as $N\rightarrow \infty $, and $a_{N}\asymp b_{N}$ means $a_{N}/b_{N}$ bounded and bounded away from zero for all $N$. For any $a,b\in \mathbb{R}$, denote %$a\wedge b=\min (a,b)$ and 
$a\vee b=\max (a,b)$. 

\subsection{Uniform Bahadur Representations}
In this section, we establish the uniform linear (Bahadur) representations of our estimators $\widehat{g}_h, \widehat{g'}_h$ and $\widehat{\tau}$ that admit the influence function expansions.

The following conditions are required and maintained throughout the remainder of the paper. %Let $s \in [2, \infty)$ be a constant integer.

\begin{assumption}\label{as:g0} 
	The function $g(t)$ is twice continuously differentiable in $t\in \mathcal{T}$.
\end{assumption}

\begin{assumption}
{\normalsize \label{ass:ciid} $\{T_{i},\boldsymbol{X}_{i},Y_{i}\}_{i=1}^{N}$
are observations drawn independently from the distribution of $(T,%
\boldsymbol{X},Y)$. The support $\mathcal{T}\times \mathcal{X}$ of $(T,
\boldsymbol{X})$ is compact. }
\end{assumption}

\begin{assumption}
\label{ass:cbounded} There exist two positive constants $%
\underline{\eta }$ and $\overline{\eta }$ such that $0<\underline{\eta }\leq
\pi(t,\boldsymbol{x})\leq \overline{\eta }<\infty $ for all $(t, 
\boldsymbol{x})\in \mathcal{T}\times \mathcal{X}$. 
\end{assumption}

\begin{assumption}
{\normalsize \label{ass:cpi0-pi*} There exist $\boldsymbol{\lambda }_{K}\in \mathbb{R}^{K}$ and a positive constant $\omega
>0$ such that $\sup_{(t,\boldsymbol{x})\in \mathcal{T}\times \mathcal{X}
}|\pi(t,\boldsymbol{x}) -\boldsymbol{\lambda }_{K}^{\top }u_{K}(t,\boldsymbol{x})|=O(K^{-\omega })$. }
\end{assumption}

\begin{assumption}
\label{as:eigen} The eigenvalues of $\mathbb{E}[u_{K}(T, 
\boldsymbol{X})u_{K}^{\top }(T,\boldsymbol{X})]$ are bounded from above and
away from zero uniformly in $K$. 
\end{assumption}

\begin{assumption}
\label{ass:regularity} \
\begin{itemize} 
\item[(i)]$\phi (a|t):=\mathbb{E}\{\mathcal{L}(Y-a)|T=t\}$ and $\widetilde{\phi }(a|t):=\mathbb{E}\{\pi(T,\boldsymbol{X})\mathcal{L}(Y-a)|T=t\}$ are
twice continuously differentiable with respect to $a,$ and all derivatives
are continuous in $t\in \mathcal{T}$ for every fixed $%
a $. Moreover, $\inf_{t\in \mathcal{T}}\partial
_{a}^{2}\widetilde{\phi }(g(t)|t)>0$.

\item[(ii)] The kernel function $\mathcal{K}(v)\geq 0$ is continuous and bounded, and $\kappa _{j1}:=\int_{-\infty }^{\infty }v^{j}\mathcal{K}
(v)dv$ satisfies $\kappa _{0,1}=1$, $\kappa _{1,1}=0$, and $\kappa
_{j,1}<\infty $ for $j=2,3,4$. 

\item[(iii)]  The density function $f_{T}(t)$ is continuous
and satisfies $\inf_{t\in \mathcal{T}}f_{T}(t)>0$.

\item[(iv)] {\normalsize The conditional density function $f_{Y|T}(y|t)$
is continuous in $t$ for every $y\in\mathbb{R}$. There exists a positive
constant $\varepsilon>0$ and a positive function $G(y|t)$ such that $%
\sup_{\|t^{\prime }-t\|\le \epsilon}f_{Y|{T}
}(y|t^{\prime })\le G(y|t)$ holds for almost all $y$, and that 
\begin{equation*}
\int \left\{\mathcal{L}(y-\delta) - \mathcal{L}(y) + \mathcal{L}^{\prime
}(y)\cdot \delta \right\}^2 G(y|t) dy = O(\delta^4)\ \text{as}\
\delta\rightarrow 0
\end{equation*}
holds uniformly over $t\in\mathcal{T}$. }

\item[(v)] The functions  $\partial _{a}^{2}\phi
(g(t)|t)$ and $e(t) := \partial _{a}^{2}\widetilde{\phi }(g(t)|t)f_{T}(t)$ are continuously differentiable in $t$,
with derivatives uniformly bounded over $t\in \mathcal{T}$.  
\end{itemize}
\end{assumption}

\begin{assumption}
{\normalsize \label{ass:vctype} } \

\begin{itemize} 
\item[(i)] {\normalsize There exists a sufficiently large constant $M$,
such that 
\begin{equation*}
\sup_{t\in \mathcal{T}}\{|g(t)-t\cdot g'(t)|\vee |g'(t)|\}\leq M.
\end{equation*}%
The function classes $\mathcal{F}_{1}=\{(y,t)\mapsto \mathcal{L}(y-\alpha
-\beta \cdot t):\alpha ,\beta \in [-M,M]\}$ and $\mathcal{F}%
_{2}=\{(y,t)\mapsto \mathcal{L}^{\prime }(y-\alpha -\beta \cdot t):\alpha, \beta \in [-M,M]\}$ are of  Vapnik-Chervonenkis (VC) type in the sense that, for some constants $A$
and $v$, 
\begin{equation*}
\sup_{Q}\mathcal{N}(\mathcal{F}_{j},\Vert \cdot \Vert _{Q,2},\varepsilon \Vert
F_{j}\Vert _{Q,2})\leq \left( \frac{A}{\varepsilon }\right) ^{v},\ j=1,2,
\end{equation*}%
where $\mathcal{N}(\mathcal{F}_{j},\Vert \cdot \Vert _{Q,2},\varepsilon )$ is the covering number of $\mathcal{F}_{j}$ at scale $\varepsilon $ with respect to the norm $\Vert \cdot \Vert _{Q,2}$, $F_{j}$ is an envelope function for $\mathcal{F}_{j},$ , and the supremum of $Q$ is taken over all finitely discrete distributions on $\mathbb{R}^{2}$. }

\item[(ii)] {\normalsize $\mathbb{E}\left[\sup_{\alpha,\beta\in[-M,M]} | 
\mathcal{L}^{\prime}(Y-\alpha-\beta\cdot T)|^q\right]<\infty$ for some $q>2$%
. }
\end{itemize}
\end{assumption}

\begin{assumption}
 \label{ass:h} \
 \begin{itemize}
\item [(i)]	Suppose that $h\asymp N^{-c_{1}}$ and $K\asymp N^{c_{2}}$ hold for some
constants $c_{1},c_{2}>0$. 
\item [(ii)]With $\zeta (K)=\sup_{(t,\boldsymbol{x})\in 
	\mathcal{T}\times \mathcal{X}}\Vert u_{K}(t,\boldsymbol{x})\Vert $,  $\zeta(K)^2K\prec N^{1/2}\prec
K^\omega$ and $\{\zeta(K)^2K^{1-2\omega}\}\vee (K^2/N)\prec
N^{1-2/q}h^3$, where $\omega$  is defined in Assumption~\ref{ass:cpi0-pi*};
\item [(iii)] $h^2\prec (Nh)^{-1/2}$. 
 \end{itemize}
\end{assumption}

Assumption \ref{ass:ciid} restricts the random variables to be bounded. This condition is restrictive but convenient for large-sample derivations. 
Assumption \ref{ass:cbounded} requires the weight function to be bounded and bounded away from zero. This condition is commonly imposed in the covariate-balancing literature. 
%Assumption \ref{as:rho} permits a wide class of estimators, including the exponential tilting, $\rho (v)=-\exp(-v-1)$, as a special case. 
Assumption \ref{ass:cpi0-pi*} requires the sieve approximation errors to shrink to zero at polynomial rates. 
Assumption \ref{as:eigen} rules out near multicollinearity in the approximating basis
functions. This condition is familiar in the sieve regression literature \citep{chen2007large}.    
Assumption \ref{ass:regularity} is similar to Condition A of \cite{fan1994robust}. Assumption \ref{ass:vctype} is needed for establishing the uniform inference results using Corollary 5.1 of \cite{chernozhukov2014gaussian}, and it is satisfied by
commonly used loss function, see Appendix \ref{app:VC} for the verification of Assumption \ref{ass:vctype}(i) under $\mathcal{L}(v)=v^2$ and $\mathcal{L}(v)=v\cdot\{\tau-\mathds{1}(v\leq 0)\}$. Assumption \ref{ass:h} is an under-smoothing condition. $\zeta (K)=O(K)$ for the power series and $\zeta (K)=O(\sqrt{K})$ for the B-splines.

Under these conditions, we establish the following asymptotic representation for $\widehat{g}, \widehat{g'}$ and their bootstraps,  $\widehat{g}^{(b)}, \widehat{g'}^{(b)}$, for $b=1,\ldots,B$. To aid the representation of the theorems, we define
\begin{align*}
& \psi _{0,h}(T_{i},\boldsymbol{X}_{i},Y_{i};t)=\frac{1}{%
e(t)}\cdot \bigg[\pi(T_{i},\boldsymbol{X}_{i})\mathcal{L}^{\prime
}(Y_{i}-g(t)-g'(t)(T_{i}-t)) \\
& \qquad \qquad \qquad -\mathbb{E}\{\pi(T_{i},\boldsymbol{X}_{i})%
\mathcal{L}^{\prime }(Y_{i}-g(t)-g'(t)(T_{i}-t))|T_{i},\boldsymbol{X}_{i}\}\bigg]\mathcal{K}_{h}(T_{i}-t) \\
& \text{and }\psi _{1,h}(T_{i},\boldsymbol{X}_{i},Y_{i};t)=%
\frac{1}{e(t)\cdot \kappa _{21}}\cdot \bigg[\pi(T_{i},\boldsymbol{X}_{i})%
\mathcal{L}^{\prime }(Y_{i}-g(t)-g'(t)(T_{i}-t)) \\
& \qquad \qquad -\mathbb{E}\{\pi(T_{i},\boldsymbol{X}_{i})%
\mathcal{L}^{\prime }(Y_{i}-g(t)-g'(t)(T_{i}-t))|T_{i},\boldsymbol{X}_{i}\}\bigg](T_{i}-t)\mathcal{K}%
_{h}(T_{i}-t)\,.
\end{align*}

\begin{theorem}
\label{thm:distribution} Under Assumptions \ref{as:TYindep}-\ref{ass:h}, we obtain the linear (Bahadur) representations for $\widehat{g}_h(t)$
and $\widehat{g'}_h(t)$ uniformly over $t$:  
\begin{equation*}
\sqrt{Nh}\left\{ \widehat{g}_h(t)-g(t)\right\} =\frac{1}{\sqrt{%
Nh}}\sum_{i=1}^{N}\psi _{0,h}(T_{i},\boldsymbol{X}%
_{i},Y_{i};t)+R_{0,N}(t)
\end{equation*}%
and 
\begin{equation*}
\sqrt{Nh^3}\left\{ \widehat{g'}_h(t)-g'(t)\right\} =\frac{1}{\sqrt{Nh^3}}\sum_{i=1}^{N}\psi _{1,%
h}(T_{i},\boldsymbol{X}_{i},Y_{i};t)+R_{1,N}(t)\,.
\end{equation*}
For $\widehat{g}^{(b)}$ and $\widehat{g'}^{(b)}$, for $b=1,\ldots,B$, we have:
\begin{equation*}
	\sqrt{Nh}\left\{ \widehat{g}^{(b)}_h(t)-\widehat{g}_h(t)\right\} =\frac{1}{\sqrt{%
	Nh}}\sum_{i=1}^{N}\{\xi^{(b)}-1\}\psi _{0,h}(T_{i},\boldsymbol{X}%
	_{i},Y_{i};t)+R^{(b)}_{0,N}(t)
	\end{equation*}%
	and 
	\begin{equation*}
	\sqrt{Nh^3}\left\{ \widehat{g'}^{(b)}_h(t)-\widehat{g'}_h(t)\right\} =\frac{1}{\sqrt{Nh^3}}\sum_{i=1}^{N}\{\xi^{(b)}-1\}\psi _{1,%
	h}(T_{i},\boldsymbol{X}_{i},Y_{i};t)+R^{(b)}_{1,N}(t)\,.
\end{equation*}
Moreover, we show that 
\begin{equation*}
\sup_{t\in \mathcal{T}}|R_{0,N}(t)|=o_{P}\left( \log^{-1} N\right)\,, \quad \sup_{t\in \mathcal{T}}|R_{1,N}(t)|=o_{P}\left( \log^{-1} N\right)\,;
\end{equation*}
and for $b=1,\ldots,B$,
$$
\sup_{t\in \mathcal{T}}|R^{(b)}_{0,N}(t)|=o_{P}\left( \log^{-1} N\right)\,, \quad  \sup_{t\in \mathcal{T}}|R^{(b)}_{1,N}(t)|=o_{P}\left( \log^{-1} N\right)\,.
$$
\end{theorem}

\begin{rk}\label{rk:bias}
In the proof of Theorem \ref{thm:distribution}, we show that the rate of the bias of of the estimators $\widehat{g}_h(t)$ and  $\widehat{g'}_h(t)$ are of $h^2$ uniformly over $t\in\mathcal{T}$. The variances $Var\left\{ \psi _{0,h}(T_{i},\boldsymbol{X}_{i},Y_{i};t)\right\}\asymp h$ and  $Var\left\{\psi _{1,h}(T_{i},\boldsymbol{X}_{i},Y_{i};t)\right\} \asymp h^3$. For the estimators $\widehat{g}_h(t)$ and  $\widehat{g'}_h(t)$, the bias is thus negligible compared to the variance, under the under-smoothing condition Assumption \ref{ass:h} (ii).
\end{rk}

The asymptotic distribution of $\widehat{\tau }(t_{1},t_{0})$, the estimator of the general continuous treatment-effect function when the treatment level changes discretely from $t_{0}$ to $t_{1}$, follows from applying Theorem \ref{thm:distribution}. 
\begin{cor}\label{cor:distribution} 
Under Assumptions \ref{as:TYindep}-\ref{ass:h}, we obtain 
\begin{align*}
& \sqrt{Nh}\left\{ \widehat{\tau }(t_{1},t_{0})-\tau(t_{1},t_{0})\right\} \\
=& \frac{1}{\sqrt{Nh}}\sum_{i=1}^{N}\{\psi _{0,h}(T_{i},%
\boldsymbol{X}_{i},Y_{i};t_{1})-\psi _{0,h}(T_{i},\boldsymbol{%
X}_{i},Y_{i};t_{0})\}+R_{1,N}(t_{1},t_{0}),
\end{align*}
and for $b=1,\ldots,B$,
\begin{align*}
& \sqrt{Nh}\left\{ \widehat{\tau}^{(b)}(t_{1},t_{0})-\widehat{\tau}(t_{1},t_{0})\right\} \\
=& \frac{1}{\sqrt{Nh}}\sum_{i=1}^{N}\{\xi^{(b)}-1\}\{\psi _{0,h}(T_{i},\boldsymbol{X}_{i},Y_{i};t_{1})-\psi _{0,h}(T_{i},\boldsymbol{X}_{i},Y_{i};t_{0})\}+R^{(b)}_{1,N}(t_{1},t_{0}),
\end{align*}
where 
\begin{equation*}
\sup_{(t_{0},t_{1})\in \mathcal{T}\times \mathcal{T}
}|R_{N}(t_{0},t_{1})|=o_{P}\left( \log^{-1} N\right) \ \text{and}\ \sup_{(t_{0},t_{1})\in \mathcal{T}\times \mathcal{T}
}|R^{(b)}_{N}(t_{0},t_{1})|=o_{P}\left(\log^{-1} N\right).
\end{equation*}
\end{cor}

\subsection{Validity of Uniform Confidence Bands}
From Theorem~\ref{thm:distribution} and Corollary~\ref{cor:distribution}, we have the variances of the asymptotic dominated terms of $\widehat{g}_h(t)-g(t)$, $\widehat{g'}_h(t)-g'(t)$ and $\widehat{\tau}(t_1,t_0)-\tau(t_1,t_0)$ are 
\begin{equation*}
\sigma _{g,N}^{2}(t,h)=\frac{Var\left\{ \psi _{0,h}(T_{i},\boldsymbol{X}_{i},Y_{i};t)\right\}}{Nh^2}\,, \quad  \sigma
_{g',N}^{2}(t)=\frac{Var\left\{\psi _{1,h}(T_{i},\boldsymbol{X}_{i},Y_{i};t)\right\}}{Nh^6}\,,
\end{equation*}
and 
\begin{equation*}
{\sigma}^2_{\tau,N}(t_{1},t_{0},h):=\frac{Var\left\{\psi _{0,h}(T,\boldsymbol{X},Y;t_{1})-\psi _{0,h}(T,\boldsymbol{X},Y;t_{0})\right\}}{Nh^2}\,,
\end{equation*}
respectively.

Applying the Lyapunov central limit theorem (CLT), Theorem~\ref{thm:distribution} and Corollary~\ref{cor:distribution}, we obtain the pointwise convergence in distribution,
\begin{align}\label{eq:gpointboot}
\lim _{N\to \infty}	\frac{\widehat{\theta}^{(b)}(\boldsymbol{w})-\widehat{\theta}(\boldsymbol{w})}{\sigma_{\theta,N}(\boldsymbol{w},h)}\overset{d}{=} \lim _{N\to \infty}	\frac{\widehat{\theta}(\boldsymbol{w})-\theta(\boldsymbol{w})}{\sigma_{\theta,N}(\boldsymbol{w},h)} \overset{d}{=}  \mathcal{N}(0,1)\,,
\end{align}
for the generic parameter notation $\theta(\boldsymbol{w}) = g(t), g'(t)$ or $\tau(t_1,t_0)$, for every $t,t_0,t_1\in \mathcal{T}$. 

Note that \eqref{eq:gpointboot} underpins the validity of pointwise inference for $g(t)$, $g'(t)$ and $\tau(t_1,t_0)$ based on the bootstrap procedure. However,  the results of pointwise convergence  in distribution  stated above cannot  be strengthened to (uniform) weak convergence in stochastic processes, because the stochastic processes $\{\widehat{g}_h(t):t\in\mathcal{T}\}$, $\{\widehat{g'}_h(t):t\in\mathcal{T}\}$, and $\{\widehat{\tau}(t_1,t_0):(t_1,t_0)\in\mathcal{T}\times \mathcal{T}\}$  are not asymptotically tight, i.e.
 \begin{align*}
 	\overline{\lim_{N\to \infty}} \sup_{\boldsymbol{w}\in\mathcal{W}}	\frac{\widehat{\theta}(\boldsymbol{w})-\theta(\boldsymbol{w})}{\sigma
 		_{\theta,N}(\boldsymbol{w},h)}\to \infty,
 \end{align*} 
with probability approaching one for  $\theta(\boldsymbol{w}) = g(t), g'(t)$ or $\tau(t_1,t_0)$.

Despite the absence of limiting distributions for the supremum of normalized processes, using the spirit of \cite{chernozhukov2014anti} and \cite{chernozhukov2014gaussian}, we shall show there exists a tight Gaussian process, whose supermum closely approximates the supremum of the normalized processes. Thereby, we can validate our uniform inference. To complete this step, we require the following additional conditions:

\begin{assumption}\label{as:sigmahatrate}
	For each of $\widehat{\sigma}_{g,N}, \widehat{\sigma}_{g',N}$ and $\widehat{\sigma}_{\tau,N}$, for every $\epsilon>0$, there exists $N_{\epsilon}$ such that for every $N\geq N_{\epsilon}$, 
	$$
	\mathbb{P}\left(\sup_{t}\left|\frac{\widehat{\sigma}_{g,N}(t,h)}{\sigma_{g,N}(t,h)}-1\right|\geq \epsilon/\log N \right)\leq \epsilon\,, \quad \mathbb{P}\left(\sup_{t}\left|\frac{\widehat{\sigma}_{g',N}(t)}{\sigma_{g',N}(t)}-1\right|\geq \epsilon/\log N\right)\leq \epsilon\,, 
	$$
	and
	$$
	\mathbb{P}\left(\sup_{t_1,t_0}\left|\frac{\widehat{\sigma}_{\tau,N}(t_1,t_0)}{\sigma_{\tau,N}(t_1,t_0)}-1\right|\geq \epsilon/\log N\right)\leq \epsilon\,.
	$$
\end{assumption}

\begin{assumption}\label{as:B1'}
	$\mathbb{E}\left[\sup_{\alpha,\beta\in[-M,M]} | 
	\mathcal{L}^{\prime}(Y-\alpha-\beta\cdot T)|^q\right]<\infty$ for some $q\geq 4$ and $\sup_{t\in\mathcal{T}}\mathbb{E}\left[\sup_{\alpha,\beta\in[-M,M]} | 
	\mathcal{L}^{\prime}(Y-\alpha-\beta\cdot T)|^4|T=t\right]<\infty$
\end{assumption}

The asymptotic validity of the proposed confidence bands can be established with any estimator of the asymptotic variances satisfying Assumption~\ref{as:sigmahatrate}. For example, as discussed in \citealp[][Remark~3.2]{chernozhukov2013inference}, the bootstrap normalized interquartile range estimator and bootstrap standard deviation, proposed in Section \ref{sec:ourinferenceframework}, satisfy the assumption under some additional conditions. Specifically, one can verify that for the bootstrap normalized interquartile range estimator, $\widehat{\sigma}^2_{g,N}(t)$ (resp. $\widehat{\sigma}^2_{g',N}(t)$ and $\widehat{\sigma}^2_{\tau,N}(t_1,t_0)$), if the corresponding target variance, ${\sigma}^2_{g,N}(t)$ (resp. $\sigma^2_{g',N}(t)$ and $\sigma^2_{\tau,N}(t_1,t_0)$) is bounded away from zero for $t\in \mathcal{T}$ (resp.  $t\in \mathcal{T}$ and  $(t_1,t_0)\in \mathcal{T}\times \mathcal{T}$). The bootstrap standard deviation satisfies the assumption when $\{Nh(\widehat{g}^{(b)}_h(t)-\widehat{g}_h(t))^2: t\in\mathcal{T}\}$ (resp. $\{Nh^3(\widehat{g'}^{(b)}_h(t)-\widehat{g'}_h(t))^2: t\in\mathcal{T}\}$) is uniformly integrable (see \citealp{kato2011Note}). Our simulation studies utilize the bootstrap standard deviation, demonstrating effective results. Furthermore, Assumption~\ref{as:B1'}, required in \cite{chernozhukov2014gaussian}, is crucial for the tight Gaussian process approximation of the estimators' asymptotic processes.

\begin{theorem}\label{thm:CI} Suppose that Assumptions~\ref{as:TYindep}-\ref{as:B1'} hold. We have 
\begin{equation*}
\lim_{N\rightarrow \infty}\mathbb{P}\left( g\in \widehat{I}_{g,N}\right)=\lim_{N\rightarrow \infty }\mathbb{P}\left( \tau_{g'}\in \widehat{I}_{g',N}\right) =\lim_{N\rightarrow \infty }\mathbb{P}%
\left( \tau\in \widehat{I}_{\tau,N}\right) =1-\alpha.
\end{equation*}
\end{theorem}

\section{Practical Details}\label{sec:NumericalDetails}
\subsection{Basis functions} \label{sec:basis}
 In our numerical studies, we set the sieve basis $u_K$ to be the product power series. Specifically, we define $u_K(T,\boldsymbol{X}) = w_{K_1}(T)\otimes w_{K_2}(\boldsymbol{X})$ and $K=(K_1+1)\cdot (K_2+1)$, where, for any positive integers $p$ and $r$, $w_p(\cdot): \mathbb{R}^r \mapsto \mathbb{R}^{pr+1}$:
$$
w_{p}(\bs{v}) = (1, \bs{v}_1^{1:p}, \ldots, \bs{v}_{r}^{1:p})^\top\,,
$$
for $j=1,\ldots, r$, $\bs{v}_j^{1:p} = ({v}_j, {v}_j^2, \ldots, {v}_j^p)$ and   ${v}_j$ is the $j^{th}$ element of any variable $\bs{v}\in\mathbb{R}^{r}$. Furthermore, the kernel function $\mathcal{K}$ is chosen as the standard normal density function.

\subsection{Data-driven tuning parameters and confidence bands}
 \label{sec:tuning} The tuning parameters $K$ and $h$ must satisfy some sufficient conditions, Assumption~\ref{ass:h}, where $h$ is required to decrease at a faster rate (i.e., undersmoothing; see Remark~\ref{rk:bias}), for the asymptotic derivation. However, those conditions do not give a unique selection in practice. This section presents two data-driven approaches for practice inference. We first use cross-validation (CV) to select the number of sieves and the pilot bandwidth that yield optimal mean-squared error estimation results.
 Then we introduce a method for selecting a smaller bandwidth that satisfies the undersmoothing condition, adapting the idea of \citet{bissantz2007non}, and an alternative approach based on the Lepski method, which controls the bias to ensure valid confidence bands \citep[see, e.g.,][]{chernozhukov2014anti, chen2025adaptive}. We also establish the validity of the Lepski method.

\subsubsection{Pilot parameters}
Specifically, in the first step, we randomly split the dataset into $F$ sets for an integer $2\leq F\leq N$. In principle, we choose the pilot tuning parameters for $g(\cdot)$, $\widehat{K} = (\widehat{K}_1+1)\cdot (\widehat{K}_2+1)$ and $\widetilde{h}$, to minimize an adjusted $F$-fold CV criteria:
\begin{equation*}
	G(K_1,K_2,h)=\sum_{j=1}^{F}\frac{1}{|S_{j}|}\sum_{k\in S_{j}}%
	\widehat{\pi }_{K}(T_{k},\mathbf{X}_{k})\mathcal{L}\{Y_{k}-\widehat{g}_{h}^{(-j)}(T_{k})\}\bigg/(1-K/N)^2\,,
\end{equation*}
over $(K_1,K_2,h)\in \{1,\ldots, P\}^2 \times \mathcal{H}$, for some positive integer $P$ and a set $\mathcal{H}$, where, for $j=1,\ldots,F$, $S_{j}$ denotes the $j^{th}$ set of the dataset, $|S_{j}|$ is the sample size of the $j$th set, and $\widehat{g}_{h}^{(-j)}$ is computed in the same way as $\widehat{g}_h$ but excluding the observations in $S_{j}$. By doing so, we perform cross-validation for choosing $h$. To avoid overfitting due to being too large $K$, we adjust the $F$-fold CV criteria by a factor $1/(1-K/N)^2$. While it is possible to conduct a full cross-validation for both $K$ and $h$, this approach is computationally intensive and may not be practical in all scenarios. In our numerical studies, we take $F=5$.
%Remark \ref{rk:bias} suggests that the rate of the optimal bias-variance trade-off bandwidth, $h\asymp N^{-1/5}$, for $\widehat{g}$. We thus recommend setting $\mathcal{H} = \text{sd}(T)\cdot N^{-1/5}\cdot[\log(N)^{-1},\log(N)]$where $\text{sd}(T)$ denotes the sample standard deviation of $\{T_i\}_{i=1}^N$. 

\subsubsection{Method~1 (Undersmoothing): Data-driven undersmooth bandwidth}
To ensure we obtain a proper undersmoothing bandwidth in the second step, \citep{bissantz2007non} suggest to use a slightly oversmoothing bandwidth as the pilot bandwidth, i.e., the pilot bandwidth $h_0 = \gamma \widetilde{h}$, for some $\gamma>1$. For the inference for $g$ and $\tau$, we set $\gamma = 1.1$.  

In the second step, we aim at choosing a positive constant $c<1$ such that $h = c \cdot h_0$ satisfies the undersmoothing requirement. Let $d_{L_\infty}(g_1,g_2)$ be the $L_\infty$ distance between two functions $g_1$ and $g_2$. %As argued in \cite{bissantz2007non}, for slightly oversmoothing bandwidths and the optimal bandwidth, $d_{L_\infty}(\widehat{g},g)$, the distance between the estimated function $\widehat{g}$ and the true function $g$ changes moderately as the bandwidth decreases, as the extrema of the estimated function get smoothed out. In contrast, for undersmoothing bandwidths, the bias is dominated by the variance, decreasing bandwidth leads to strongly increasing artificial oscilations. 
Observing the rates of the bias and variance of the estimators in Remark~\ref{rk:bias}, the supremum distances, $d_{L_\infty}(\widehat{g},g):=\sup_{t\in\mathbb{G}}|\widehat{g}_h(t)-g(t)|$, for a suitably fine grid $\mathbb{G}\in \mathcal{T}$, remain relatively stable for bandwidths close to the optimal bias-variance trade-off bandwidth. However, for undersmoothing bandwidths such that the bias is negligible compared to the variance, these distances increase sharply as the bandwidth decreases. Thus, with the bandwidth decreases from a slightly oversmoothing scale, we should observe a sudden steep increase in  $d_{L_\infty}(\widehat{g},g)$. While we can not compute $d_{L_\infty}(\widehat{g},g)$, \cite{bissantz2007non} proposed to calculate the $L_\infty$-distance between $\widehat{g}$ with decreasing bandwidths to find the sudden steep increase.

Specifically, consider an integer $J$; we compute the estimator with bandwidth $h_j = (J-j)/J\cdot h_0$, denoted by $\widehat{g}_{h_j}$, for $j=0,\ldots, J-1$. Then we compute $d_{L_{\infty}}(\widehat{g}_{h_j},\widehat{g}_{h_{j-1}})$ for $j=1,\ldots,J-1$. We choose the undersmoothing bandwidth to be $\widehat{h}_{\text{u}}:=(J-j)/J\cdot h_0$ for the smallest $j$ where we observe a sudden steep increase (a turning point) in the $d_{L_{\infty}}(\widehat{g}_{h_j},\widehat{g}_{h_{j-1}})$'s, as shown in Figure~\ref{fig:USCampaign} in our empirical study in Section~\ref{sec:empirics}. For the derivative $g'$, Remark~\ref{rk:bias} suggests the optimal rate of bandwidth is of a larger rate than that for $f$. We take $\widehat{K}$ the same as for $g$, but the pilot bandwidth $h_0 = 4\widetilde{h}\cdot N^{1/5}\cdot N^{-1/7}$. Then we conduct the undersmoothing procedure as that for the inference of $g$. The resulting undersmoothing bandwidth is denoted as $\widehat{h^{\prime}}_{\text{u}}$.

As recommended by \cite{bissantz2007non}, we take $J=20$. In our simulation studies, we consider the conditional average and quantile dose-response functions, i.e., taking $\mathcal{L}(v)=v^2$ and $\mathcal{L}(v) = v\{q - \mathbbm{1}(v\leq 0)\}$ for some $q\in(0,1)$, respectively. For all the cases, we found that the significant sudden steep increase mostly corresponds to $j \in [15, 18]$. For simplicity, we set $j = 18$ throughout the simulation studies.

\subsubsection{Method~2 (Lepski): Data-driven confidence bands}
We adapt the Lepski method \citep[see e.g.,][]{chernozhukov2014anti, chen2025adaptive} to our context. %In this section, for clearance, we rewrite our local polynomial estimators $\widehat{g}$ and $\widehat{g^{\prime}}$ with bandwidth $h$ as $\widehat{g}_h$ and $\widehat{g^{\prime}}_h$, respectively. 
Let ${J}_{\max,N}$ and ${J}_{\min,N}$ be two sequences of integers, and the corresponding candidate set of bandwidths be $\mathcal{H}_N:=\{h=2^{-\ell}: \ell\in [J_{\min,N},J_{\max,N}]\cap \mathbb{Z}\}$. For any bandwidth $h\in\mathcal{H}_N$, let $D_{h}(t) = \widehat{g}_{h}(t) - g(t)$. Defining $\mathcal{V}_N := \{(h,h_2)\in\mathcal{H}_N: h_2<h\}$, for any $h, h_2 \in \mathcal{V}_N$, we have $D_{h}(t) - D_{h_2}(t) = \widehat{g}_{h}(t) - \widehat{g}_{h_2}(t)$. Moreover, the weighted bootstrap version of $D_{h}(t) - D_{h_2}(t)$ is 
$$
D^{(b)}_{h}(t) - D^{(b)}_{h_2}(t) := \{\widehat{g}^{(b)}_{h}(t) - \widehat{g}_{h}(t)\} - \{\widehat{g}^{(b)}_{h_2}(t) - \widehat{g}_{h_2}(t)\}\,,
$$
and 
$$
Z(t,h,h_2):= \frac{D^{(b)}_{h}(t) - D^{(b)}_{h_2}(t)}{\widehat{\sigma}_{g,N}(t,h,h_2)}
$$
is a stochastic process, where $\widehat{\sigma}_{g,N}(t,h,h_2)$ is a consistent estimator of $\sigma_{g,N}(t,h,h_2)$, the truncated asymptotic standard deviation of $D^{(b)}_{h}(t) - D^{(b)}_{h_2}(t)$. Specifically,
\[
\sigma_{g,N}^{2}(t,h,h_2):=\widetilde{\sigma}_{g,N}^{2}(t,h,h_2)\vee\left\{ c_{\sigma}^{2}\sigma_{g,N}^{2}(t,h_2)\right\} ,
\]
where
\[
\widetilde{\sigma}_{g,N}^{2}(t,h,h_2):=Var\left\{ \frac{1}{Nh}\sum_{i=1}^{N}\varphi_{0,h}(T_{i},\boldsymbol{X}_{i},Y_{i};t)-\frac{1}{Nh_2}\sum_{i=1}^{N}\varphi_{0,h_2}(T_{i},\boldsymbol{X}_{i},Y_{i};t)\right\}\,,
\]
the small constant $c_{\sigma}$ is chosen so that with probability
approaching one $\widetilde{\sigma}_{g,N}^{2}(t,h,h_2)>c_{\sigma}^{2}\sigma_{g,N}^{2}(t,h_2)$
for $h>\varpi h_2$ and $\varpi$ is a constant (see Lemma
\ref{lem:bound on the variance} in the supplementary material). 
In our numerical studies, we take $\widehat{\sigma}_{g,N}(t,h,h_2)$ to be the sample standard deviation of $\{D^{(b)}_{h}(t) - D^{(b)}_{h_2}(t)\}_{b=1}^B$ and $c_{\sigma}=0.25$.

Let $\gamma_N$ be a sequence of positive numbers converging to zero. We define $\widetilde{c}_N(\gamma_N)$ be the conditional $(1-\gamma_N)$-quantile of $\sup_{t\in\mathcal{T}, (h,h_2)\in\mathcal{V}_N}|Z(t,h,h_2)|$ given the observed data $\{(T_i,\boldsymbol{X}_i,Y_i):1\leq i \leq N\}$. Letting $\mathcal{H}_{N,h}:=\{h_2\in\mathcal{H}_N: h_2 < h\}$, for some constant $v>1$, which is independent of $N$, a Lepski-type estimator of the bandwidth is
\begin{equation}\label{def:h_hat}
\widehat{h}:=\sup\left\{h\in\mathcal{H}_N: \sup_{h_2\in\mathcal{H}_{N,h}}\sup_{t\in\mathcal{T}}\left|\frac{\widehat{g}_{h}(t) - \widehat{g}_{h_2}(t)}{\widehat{\sigma}_{g,N}(t,h,h_2)}\right|\leq v \widetilde{c}_N(\gamma_N)\right\}\,.
\end{equation}
The Lepski-type data-driven confidence interval further corrects the confidence interval $I_{g,N}(t)$ for the bias introduced by the estimated bandwidth $\widehat{h}$. Specifically, the final Lepski-type confidence interval is defined as 
$$
\tilde{I}_{g,N}:= [\widehat{g}_{\widehat{h}}(t) - \{\widehat{c}_N(\alpha)+u_N\widetilde{c}_N(\gamma_N)\}\cdot\widehat{\sigma}_{g,N}(t,\widehat{h}), \widehat{g}_{\widehat{h}}(t) + \{\widehat{c}_N(\alpha)+u_N\widetilde{c}_N(\gamma_N)\}\cdot \widehat{\sigma}_{g,N}(t,\widehat{h})]\,,
$$
for some adjustment parameter $u_{N}$, where $\widehat{c}_{N}(\alpha)$ is the conditional $(1-\alpha)$-quantile
of $\sup_{t\in\mathcal{T},h\in\mathcal{H}_{N}}\left|\left(\widehat{g}_{h}^{(b)}(t)-\widehat{g}_{h}(t)\right)/\widehat{\sigma}_{g,N}(t,h)\right|$ given the observed data $\{(T_i,\bs{X}_i,Y_i): 1\leq i \leq N\}$.

Following \cite{chernozhukov2014anti} and \cite{chen2025adaptive}, we set $J_{\max,N}= \lceil \max(\tilde{J}_N, -\log_2(\tilde{h}_0/10)) \rceil$ and $J_{\min,N}=\lceil\max(\log(J_{\max,N}), -\log_2(\tilde{h}_0\cdot \log^{1/5}(N)/3)) \rceil$, where $\tilde{J}_N = \log_2(10\cdot N\cdot \tilde{h}_0/\log^4(N))$, $\tilde{h}_0 = \widetilde{h}$, the CV bandwidth for $g$. Moreover, $\gamma_N = \min(0.5, \{\log(J_{\max,N})/J_{\max,N}\}^{1/2})$, $v=1.1$, $u_N=0.5$.

The data-driven Lepski-type bandwidth and confidence interval for the derivative $g{'}$ can be constructed analogously by replacing $\widehat{g}_h, \widehat{g}^{(b)}_h$ with $\widehat{g}^{\prime}_h$ and $\widehat{g^{\prime}}^{(b)}_h$, respectively, for all $h$ thoroughly in the above algorithm. We denote the resulting bandwidth and confidence interval as $\widehat{h^{\prime}}$ and $\tilde{I}_{g^{\prime},N}$, respectively. Additionally, 
$\tilde{h}_0$ in $J_{\max,N}$ is replaced by $3\widetilde{h}\cdot N^{1/5}\cdot N^{-1/7}$ in our numerical studies. The resulting candidate set of bandwidths is denoted by $\mathcal{H}^{\prime}_N$.

\subsection{Validity of the Lepski method}

We define the approximation of $g(h)$ and $g^{\prime}(t)$ under
the bandwidth $h$ as 
\[
(g_{h}(t),g_{h}^{\prime}(t)):=\argmin_{(\theta_{1},\theta_{2})\in\mathbb{R}^{2}}\e\left[\pi_{0}(T,\bs X)\mathcal{L}\left\{ Y-\theta_{1}-\theta_{2}(T-t)\right\} \mathcal{K}_{h}\left(T-t\right)\right].
\]

\begin{assumption}
\label{assu:approximation error} Suppose $g$ lies in a function class $\mathcal{G}$.
There exist constants $0<h_{0},c_{3},C_{3},\underline{s},\bar{s}<\infty$
such that for every $g\in\mathcal{G}$, %there exists $s\in[\underline{s},\bar{s}]$
with
\[
c_{3}h^{2}\leq\sup_{t\in\mathcal{T}}\left|g_{h}(t)-g(t)\right|\leq C_{3}h^{2}\ \mathrm{and}\ \sup_{t\in\mathcal{T}}\left|g_{h}^{\prime}(t)-g^{\prime}(t)\right|\leq C_{3}h^{2}
\]
for all $h\leq h_{0}$.
\end{assumption}

%In fact, under Assumptions \ref{assu:approximation error}, we can define a map:
%\begin{equation}
%s:\mathcal{G}\to[\underline{s},\bar{s}],\ \ g\mapsto s(g).\label{eq:smoothness map}
%\end{equation}
%See the discussion below \citet[Condition L2]{chernozhukov2014anti}.
\begin{assumption}
\label{assu:candidate set}There exist constants $c_{4},C_{4}>0$
such that for every $g\in\mathcal{G}$ there exist $h\in\mathcal{H}_{N}$ and $h^{\prime}\in\mathcal{H}^{\prime}_{N}$
with
\[
\left(c_{4}\frac{\log N}{N}\right)^{1/5}\leq h\leq\left(C_{4}\frac{\log N}{N}\right)^{1/5}\,,
\]
and
\[
\left(c_{4}\frac{\log N}{N}\right)^{1/7}\leq h^{\prime}\leq\left(C_{4}\frac{\log N}{N}\right)^{1/7}\,.
\]
%for the map $s:g\mapsto s(g)$ defined in (\ref{eq:smoothness map}).
\end{assumption}

\begin{assumption}
\label{assu:uniform Bahadur}It holds that for every $b=0,\ldots,B$,
\[
\sqrt{Nh}\left\{ \widehat{g}_{h}^{(b)}(t)-g_{h}(t)\right\} =\frac{1}{\sqrt{Nh}}\sum_{i=1}^{N}\xi_{i}^{(b)}\psi_{0,h}(T_{i},\boldsymbol{X}_{i},Y_{i};t)+\widetilde{R}_{0,N}(t,h,g)
\]
and 
\[
\sqrt{Nh^{3}}\left\{ \widehat{g^{\prime}}_{h^{\prime}}^{(b)}(t)-g_{h^{\prime}}^{\prime}(t)\right\} =\frac{1}{\sqrt{N(h^{\prime})^{3}}}\sum_{i=1}^{N}\xi_{i}^{(b)}\psi_{1,h^{\prime}}(T_{i},\boldsymbol{X}_{i},Y_{i};t)+\widetilde{R}_{1,N}(t,h^{\prime},g)
\]
where $\xi_{i}^{(0)}\equiv1$ and $\widetilde{R}_{0,N}(t,h,g)$ and
$\widetilde{R}_{1,N}(t,h^{\prime},g)$ satisfy
\[
\sup_{g\in\mathcal{G}}\mathbb{P}\left(\sup_{t\in\mathcal{T},h\in\mathcal{H}_{N},h^{\prime} \in \mathcal{H}^{\prime}_N}\left\{ \left|\widetilde{R}_{0,N}(t,h,g)\right|\vee\left|\widetilde{R}_{1,N}(t,h^{\prime},g)\right|\right\} \geq o((\log N)^{-1/2})\right)=o(1)
\]
as $N\to\infty$.
\end{assumption}

\begin{assumption}
\label{assu:consistency of estimated var}Let $\overline{\mathcal{V}}_{N}:=\{(t,h,h_2):t\in\mathcal{T},h,h_2\in\mathcal{H}_{N},h_2<h\}$, $\overline{\mathcal{V}}^{\prime}_{N}:=\{(t,h,h_2):t\in\mathcal{T},h,h_2\in\mathcal{H}^{\prime}_{N},h_2<h\}$,
and $\delta_{\sigma,N}$ be a sequence of nonnegative numbers satisfying
$\delta_{\sigma,N}=o((\log N)^{-1})$. It holds that 
\[
\sup_{g\in\mathcal{G}}\mathbb{P}\left(\sup_{t\in\mathcal{T},h\in\mathcal{H}_{N}}\left|\frac{\widehat{\sigma}_{g,N}(t,h)}{\sigma_{g,N}(t,h)}-1\right|\geq\delta_{\sigma,N}\right)=1-o(1)\,,
\]
\[
\sup_{g\in\mathcal{G}}\mathbb{P}\left(\sup_{t\in\mathcal{T},h\in\mathcal{H}^{\prime}_{N}}\left|\frac{\widehat{\sigma}_{g,N}^{\prime}(t,h)}{\sigma_{g,N}^{\prime}(t,h)}-1\right|\geq\delta_{\sigma,N}\right)=1-o(1)\,,
\] 
\[
\sup_{g\in\mathcal{G}}\mathbb{P}\left(\sup_{(t,h,h_2)\in\overline{\mathcal{V}}_{N}}\left|\frac{\widehat{\sigma}_{g,N}(t,h,h_2)}{\sigma_{g,N}(t,h,h_2)}-1\right|\geq o((\log N)^{-1})\right)=1-o(1)\,,
\]
and 
\[
\sup_{g\in\mathcal{G}}\mathbb{P}\left(\sup_{(t,h,h_2)\in\overline{\mathcal{V}}^{\prime}_{N}}\left|\frac{\widehat{\sigma}_{g^{\prime},N}(t,h,h_2)}{\sigma_{g^{\prime},N}(t,h,h_2)}-1\right|\geq o((\log N)^{-1})\right)=1-o(1)
\]
as $N\to\infty$.
\end{assumption}

\begin{assumption}
\label{assu:conditional var is bounded below}There exists some constant
$c>0$ such that 
\[
c^{2}\leq\sigma_{Y\mid T,\bs X}^{2}:=Var\left(\mathcal{L}^{\prime}(Y_{i}-g(t)-g'(t)(T_{i}-t))\mid T,\bs X\right).
\]
\end{assumption}

\begin{assumption}
\label{assu:kernel VC}The function class $\{t\mapsto\mathcal{K}((t-t^{\prime})/h):t^{\prime}\in\mathcal{T},h\in\mathcal{H}_{N}\}$
is VC$(b,a,v)$ type for some constants $b,a$, and $v$ independent
of $N$, where the definition of VC type class can be found in \citet[P. 1801]{chernozhukov2014anti}.
\end{assumption}

Assumptions \ref{assu:approximation error} poses some restrictions
on the approximation error, which is a weak assumption similar to Condition~L2 in \cite{chernozhukov2014anti}. %The upper bound can be established from the proof of Theorem~\ref{thm:distribution} (see Remark~\ref{rk:bias}). The lower bound holds for $g$ in a dense subset of smooth function space \citep[see e.g.,][]{chernozhukov2014anti}. 
Assumption~\ref{assu:candidate set} is satisfied by the construction of $\mathcal{H}_N$ and $\mathcal{H}^{\prime}_N$. Assumption~\ref{assu:uniform Bahadur} can be established using the similar arguments for Theorem~\ref{thm:distribution}. Assumption~\ref{assu:consistency of estimated var} is similar to Assumption~\ref{as:sigmahatrate}.
Assumption~\ref{assu:kernel VC} is a mild assumption, see \citet[P. 911]{gine2002Rates}. We have the following theorem.
\begin{thm}
\label{thm:adaptive UCB}Suppose that Assumptions \ref{as:TYindep}-\ref{ass:cbounded},
\ref{ass:regularity}-\ref{ass:vctype}, \ref{as:B1'}-\ref{assu:kernel VC}
hold. In addition, we assume that $\left(\log N\right)^{4}/\left\{N(h_{\min,N}\wedge (h^{\prime}_{\min,N})^3)\right\}=o(N^{-c_{5}})$,
$h_{\max,N}, h^{\prime}_{\max,N}\to0$, $h_{\min,N}/h_{\max,N}, h^{\prime}_{\min,N}/h^{\prime}_{\max,N}\to0$, $\gamma_{N}\to0$,
$\left|\log\gamma_{N}\right|=O(\log N)$ and $\delta_{\sigma,N}u_{N}=o((\log N)^{-1})$,
where $h_{\min,N}:=\inf\mathcal{H}_{N}$, $h_{\max,N}:=\sup\mathcal{H}_{N}$, $h^{\prime}_{\min,N}:=\inf\mathcal{H}^{\prime}_{N}$, $h^{\prime}_{\max,N}:=\sup\mathcal{H}^{\prime}_{N}$
and $c_{5}>0$ is a constant. Then it holds that (i)
\[
\liminf_{N\to\infty}\inf_{g\in\mathcal{G}}\ \mathbb{P}\left(g(t)\in\tilde{I}_{g,N}(t),\ \forall t\in\mathcal{T}\right)\geq1-\alpha
\]
and (ii)
\[
\liminf_{N\to\infty}\inf_{g\in\mathcal{G}}\ \mathbb{P}\left(g^{\prime}(t)\in\tilde{I}_{g^{\prime},N}(t),\ \forall t\in\mathcal{T}\right)\geq1-\alpha.
\]
\end{thm}

\section{Numerical Studies} \label{sec:numerical} 

\subsection{Simulation Study}

 \label{sec:simulation} 
We conduct a simulation study on a continuous treatment model to assess the
finite performance of the proposed estimators $\widehat{g}$ and $\widehat{%
\partial _{t}g}$. We consider three response functions. Specifically, let $U_w$, $U_t$ and $U_y$ be three independent standard normal random variables. The models are
given by $\boldsymbol{X} = (Z,W)^\top$, where
\begin{eqnarray*}
	&&Z\sim \text{Uniform}(-0.65,0.65),~W = 0.5Z + 0.5U_w,~T=0.1W+0.1Z+0.4U_t.\\
	&&\mathbf{DGP0:} \quad  Y^*(t) = \exp(0.2Z -0.5W)+0.1U_y,\\
	&&\mathbf{DGP1L:} \quad  Y^*(t) = 0.6t+ 0.7Z -2W+0.1U_y~\text{(linear in $t$)},\\
	&&\mathbf{DGP1NL:} \quad  Y^*(t) = 0.75\exp(t)+ 0.6Z^3-3W+0.1U_y~\text{(non-linear in $t$)}.
\end{eqnarray*}%
We generate $J=500$ samples of size $N=400$, 800 and 1200, respectively. With $%
\mathcal{L}(v)=v^{2}$ and $\mathcal{L}(v)=v\{q-\mathbbm{1}(v\leq 0)\}$, for $q = 0.25, 0.35$ and $0.45$,
respectively, we compute the estimators $\widehat{g}$ and $\widehat{\partial
_{t}g}$ for the conditional average response $g(t)=\mathbb{E}\{Y^{\ast
}(t)\}$ and the conditional quantile responses $g(t)=F_{Y^{\ast
}(t)}^{-1}(q)$, over a grid of 25 equispaced points for $t \in \mathcal{V} = [q_{0.05}(T), q_{0.95}(T)]$, where $\mathcal{V}$ spans the 5th to 95th percentiles of $T$. Note that our estimators $\widehat{g}$ and $\widehat{g'}$ have closed-form solutions for the conditional average dose-response, but not for the conditional quantile dose-response function.
The numerical minimization for estimating $g(t)=F_{Y^{\ast}(t)}^{-1}(q)$, for any quantile level $q$, is time-consuming. To speed up the computation, we use the
iteratively reweighted least squares algorithm proposed by \cite%
{lejeune1988quantile}. 
Within the bootstrap procedure, we set the number of bootstrap replications to $B=500$.

We compute the naive estimators, $\widehat{g}_{\text{N}}$ and $\widehat{\partial _{t}g}_{\text{N}}$, by setting $\widehat{\pi }_{K}\equiv 1$.
Empirical coverage probabilities, with the average widths, of the two adaptive confidence bands introduced in Section~\ref{sec:NumericalDetails} for $g$ from DGP0, DGP1L and DGP1NL are reported in Tables~\ref{table:CovProb} and \ref{table:CovProb_Lepski}, at confidence levels of $0.99$, $0.95$, and $0.90$. Furthermore, Tables~\ref{table:RejProb} and \ref{table:RejProb_Lepski} present the empirical rejection probabilities of the null hypothesis $H_{0}$ in \eqref{H0:g'}, at significance levels of $0.01$, $0.05$, and $0.1$. It is important to note that while the null hypothesis $H_{0}$ in \eqref{H0:g'} holds under DGP0, it is not valid under DGP1L and DGP1NL. All these tables collectively affirm the effectiveness of the confounding adjustments within our methodology, as evident from the favorable size of the uniform inferences and the test power, with comparable widths of the confidence bands. Overall, the Lepski method gives a more conservative but narrower confidence band than the undersmoothing method. %Comparing the quantiles, the advantage of our proposed method diminishes for the extreme quantiles due to decreasing local sample sizes.

\begin{table}[t]
	\caption{Empirical coverage probability(mean width) of the confidence band for $g$ calculated from 500 Monte-Carlo simulations using Method~1 (Undersmoothing)}
	\label{table:CovProb}{\normalsize {\footnotesize \ \centering
	\resizebox{\textwidth}{!}{  
			\begin{tabular}{ccc|ccc|ccc|ccc}
				\hline
				&&&\multicolumn{3}{c}{DGP0}&    \multicolumn{3}{c}{DGP1L}&\multicolumn{3}{c}{DGP1NL}\\
				&$N$  & Method &99\%& 95\%&90\%&99\%& 95\%&90\%&99\%& 95\%&90\%\\
				\hline
				\multirow{6}{*}{Average}
				&\multirow{2}{*}{400} &  proposed& 0.970 (0.7) & 0.898 (0.5)& 0.842 (0.4)& 0.978 (4.1)& 0.926 (2.8)& 0.868 (2.4)& 0.968 (4.2)& 0.904 (2.9)& 0.860 (2.5)  \\
				&&naive& 0.944 (0.7) & 0.830 (0.5)& 0.734 (0.4)& 0.958 (3.9)& 0.836 (2.7)& 0.732 (2.4)& 0.944 (3.9)& 0.782 (2.7)& 0.684 (2.4)\\
				&\multirow{2}{*}{800}& proposed& 0.978 (0.4) & 0.940 (0.3)& 0.868 (0.3)& 0.978 (1.9)& 0.938 (1.6)& 0.872 (1.4)& 0.978 (1.9)& 0.920 (1.6)& 0.846 (1.5)\\
				&&naive& 0.922 (0.4) & 0.776 (0.3)& 0.648 (0.3)& 0.896 (2.0)& 0.748 (1.6)& 0.624 (1.5)& 0.864 (2.0)& 0.670 (1.6)& 0.502 (1.5)\\
			
                &\multirow{2}{*}{1200}& proposed& 0.978 (0.2) & 0.926 (0.2)& 0.854 (0.1)& 0.958 (0.6)& 0.888 (0.5)& 0.818 (0.5)& 0.968 (1.1)& 0.900 (0.9)& 0.820 (0.8)\\
				&&naive& 0.720 (0.2) & 0.460 (0.2)& 0.308 (0.1)& 0.656 (0.7)& 0.358 (0.6)& 0.222 (0.5)& 0.540 (1.1)& 0.278 (0.9)& 0.174 (0.9) \\
                \hline
				\multirow{6}{*}{$q = 0.45$}
				&\multirow{2}{*}{400}&proposed& 1.000 (1.7) & 0.976 (1.1)& 0.934 (0.9)& 0.992 (5.1)& 0.950 (3.2)& 0.906 (2.7)& 0.992 (8.4)& 0.968 (5.4)& 0.940 (4.4) \\
				&&naive& 0.994 (1.5) & 0.962 (1.0)& 0.914 (0.8)& 0.984 (5.3)& 0.926 (3.3)& 0.870 (2.7)& 0.986 (7.7)& 0.928 (5.0)& 0.842 (4.1) \\		&\multirow{2}{*}{800}& proposed & 0.996 (0.6) & 0.968 (0.4)& 0.940 (0.4)& 0.996 (2.2)& 0.972 (1.7)& 0.930 (1.5)& 0.998 (3.4)& 0.964 (2.5)& 0.916 (2.2)\\
				&&naive& 0.978 (0.6) & 0.900 (0.4)& 0.842 (0.4)& 0.974 (2.0)& 0.892 (1.6)& 0.784 (1.4)& 0.964 (3.2)& 0.826 (2.4)& 0.704 (2.2)\\
			
                &\multirow{2}{*}{1200}&proposed& 0.990 (0.3) & 0.932 (0.2)& 0.856 (0.2)& 0.974 (1.0)& 0.894 (0.8)& 0.806 (0.7)& 0.988 (1.6)& 0.906 (1.3)& 0.838 (1.2) \\
				&&naive & 0.908 (0.3) & 0.684 (0.2)& 0.566 (0.2)& 0.826 (1.0)& 0.560 (0.8)& 0.410 (0.7)& 0.748 (1.7)& 0.472 (1.3)& 0.346 (1.2)\\\hline
				\multirow{6}{*}{$q = 0.35$}
				&\multirow{2}{*}{400}& proposed& 0.996 (1.5) & 0.982 (0.9)& 0.954 (0.8)& 0.992 (6.1)& 0.958 (3.8)& 0.902 (3.1)& 1.000 (8.6)& 0.976 (5.5)& 0.910 (4.5) \\
				&&naive& 0.994 (1.3) & 0.962 (0.9)& 0.914 (0.7)& 0.980 (5.1)& 0.930 (3.3)& 0.866 (2.8)& 0.992 (8.9)& 0.924 (5.6)& 0.840 (4.6)\\
				&\multirow{2}{*}{800}& proposed& 0.994 (0.6) & 0.964 (0.5)& 0.918 (0.4)& 0.986 (2.3)& 0.932 (1.7)& 0.878 (1.5)& 0.978 (3.4)& 0.944 (2.6)& 0.876 (2.3) \\
				&&naive  & 0.978 (0.6) & 0.908 (0.5)& 0.836 (0.4)& 0.970 (2.2)& 0.876 (1.7)& 0.772 (1.5)& 0.958 (3.7)& 0.842 (2.8)& 0.714 (2.4) \\
			
                &\multirow{2}{*}{1200}& proposed& 0.988 (0.3) & 0.936 (0.2)& 0.860 (0.2)& 0.968 (1.0)& 0.888 (0.8)& 0.828 (0.8)& 0.970 (1.9)& 0.898 (1.5)& 0.804 (1.3) \\
				&&naive  & 0.894 (0.3) & 0.676 (0.2)& 0.528 (0.2)& 0.812 (1.1)& 0.606 (0.9)& 0.482 (0.8)& 0.786 (1.8)& 0.552 (1.4)& 0.390 (1.3)\\
                \hline
				\multirow{6}{*}{$q = 0.25$}
				&\multirow{2}{*}{400}& proposed & 0.994 (1.5) & 0.960 (1.0)& 0.906 (0.8)& 0.988 (5.7)& 0.920 (3.6)& 0.868 (3.0)& 0.988 (8.1)& 0.932 (5.3)& 0.862 (4.4) \\
				&&naive& 0.988 (1.4) & 0.944 (0.9)& 0.906 (0.7)& 0.976 (5.3)& 0.898 (3.4)& 0.816 (2.9)& 0.980 (8.3)& 0.884 (5.4)& 0.790 (4.5)\\
				&\multirow{2}{*}{800}& proposed& 0.978 (0.6) & 0.904 (0.4)& 0.830 (0.4)& 0.976 (2.4)& 0.904 (1.8)& 0.824 (1.6)& 0.968 (3.7)& 0.886 (2.8)& 0.812 (2.5) \\
				&&naive& 0.966 (0.6) & 0.866 (0.4)& 0.760 (0.4)& 0.936 (2.4)& 0.800 (1.8)& 0.710 (1.6)& 0.936 (3.6)& 0.768 (2.8)& 0.664 (2.5)\\
			
                &\multirow{2}{*}{1200}& proposed& 0.982 (0.3) & 0.912 (0.2)& 0.830 (0.2)& 0.948 (1.2)& 0.860 (0.9)& 0.766 (0.8)& 0.950 (1.9)& 0.842 (1.5)& 0.752 (1.4) \\
				&&naive& 0.902 (0.3) & 0.730 (0.2)& 0.610 (0.2)& 0.806 (1.1)& 0.578 (0.9)& 0.438 (0.8)& 0.782 (1.9)& 0.536 (1.5)& 0.406 (1.3) \\
                \hline
			\end{tabular}} }  }
	\end{table}

    \begin{table}[t]
	\caption{Empirical coverage probability(mean width) of the confidence band for $g$ calculated from 500 Monte-Carlo simulations using the Method~2 (Lepski)}
	\label{table:CovProb_Lepski}{\normalsize {\footnotesize \ \centering
	\resizebox{\textwidth}{!}{  
			\begin{tabular}{ccc|ccc|ccc|ccc}
				\hline
				&&&\multicolumn{3}{c}{DGP0}&    \multicolumn{3}{c}{DGP1L}&\multicolumn{3}{c}{DGP1NL}\\
				&$N$  & Method &99\%& 95\%&90\%&99\%& 95\%&90\%&99\%& 95\%&90\%\\
				\hline
				\multirow{6}{*}{Average}
				&\multirow{2}{*}{400} &  proposed& 0.998 (0.2) & 0.986 (0.2)& 0.980 (0.2)& 0.986 (0.9)& 0.956 (0.8)& 0.922 (0.8)& 0.984 (0.9)& 0.944 (0.8)& 0.914 (0.8)\\
				&&naive& 0.992 (0.3) & 0.960 (0.2)& 0.914 (0.2)& 0.984 (1.3)& 0.954 (1.2)& 0.914 (1.1)& 0.970 (1.3)& 0.916 (1.2)& 0.838 (1.1) \\
				&\multirow{2}{*}{800}& proposed& 0.996 (0.1) & 0.982 (0.1)& 0.976 (0.1)& 0.976 (0.7)& 0.954 (0.6)& 0.922 (0.6)& 0.976 (0.7)& 0.940 (0.6)& 0.912 (0.6)\\
				&&naive& 0.958 (0.2) & 0.892 (0.2)& 0.834 (0.1)& 0.934 (1.0)& 0.820 (0.9)& 0.730 (0.8)& 0.864 (1.0)& 0.690 (0.9)& 0.568 (0.8)\\
			
                &\multirow{2}{*}{1200}& proposed& 0.992 (0.1) & 0.974 (0.1)& 0.952 (0.1)& 0.978 (0.3)& 0.936 (0.3)& 0.910 (0.3)& 0.960 (0.5)& 0.924 (0.5)& 0.884 (0.4) \\
				&&naive& 0.846 (0.1) & 0.656 (0.1)& 0.538 (0.1)& 0.738 (0.5)& 0.532 (0.4)& 0.406 (0.4)& 0.596 (0.8)& 0.368 (0.7)& 0.236 (0.6)\\
                \hline
				\multirow{6}{*}{$q = 0.45$}
				&\multirow{2}{*}{400}&proposed& 0.996 (0.3) & 0.982 (0.2)& 0.974 (0.2)& 0.988 (0.9)& 0.972 (0.8)& 0.964 (0.7)& 0.976 (1.3)& 0.960 (1.1)& 0.938 (1.1) \\
				&&naive& 0.998 (0.3) & 0.984 (0.3)& 0.974 (0.3)& 0.996 (1.1)& 0.970 (1.0)& 0.942 (0.9)& 0.992 (1.7)& 0.964 (1.5)& 0.922 (1.4)\\				&\multirow{2}{*}{800}& proposed & 1.000 (0.2) & 0.998 (0.2)& 0.994 (0.2)& 0.998 (0.7)& 0.988 (0.6)& 0.984 (0.5)& 0.994 (1.0)& 0.990 (0.9)& 0.978 (0.8) \\
				&&naive& 0.996 (0.2) & 0.968 (0.2)& 0.950 (0.2)& 0.972 (0.8)& 0.940 (0.7)& 0.890 (0.7)& 0.958 (1.2)& 0.868 (1.1)& 0.800 (1.0)\\
			
                &\multirow{2}{*}{1200}&proposed& 1.000 (0.1) & 0.998 (0.1)& 0.992 (0.1)& 0.994 (0.5)& 0.978 (0.4)& 0.948 (0.4)& 0.986 (0.8)& 0.954 (0.7)& 0.934 (0.6) \\
				&&naive & 0.958 (0.2) & 0.870 (0.1)& 0.796 (0.1)& 0.894 (0.6)& 0.786 (0.5)& 0.686 (0.5)& 0.840 (0.9)& 0.658 (0.8)& 0.504 (0.8) \\\hline
				\multirow{6}{*}{$q = 0.35$}
				&\multirow{2}{*}{400}& proposed& 1.000 (0.3) & 0.994 (0.2)& 0.994 (0.2)& 0.998 (1.1)& 0.994 (0.9)& 0.990 (0.9)& 0.996 (1.6)& 0.984 (1.4)& 0.980 (1.3) \\
				&&naive& 0.994 (0.3) & 0.978 (0.3)& 0.970 (0.3)& 0.994 (1.3)& 0.970 (1.1)& 0.956 (1.0)& 0.986 (1.9)& 0.964 (1.7)& 0.936 (1.5)\\
				&\multirow{2}{*}{800}& proposed& 1.000 (0.2) & 0.994 (0.2)& 0.992 (0.2)& 1.000 (0.7)& 0.988 (0.6)& 0.980 (0.6)& 1.000 (1.1)& 0.980 (1.0)& 0.964 (0.9) \\
				&&naive & 0.986 (0.2) & 0.960 (0.2)& 0.938 (0.2)& 0.970 (0.9)& 0.940 (0.8)& 0.908 (0.7)& 0.946 (1.3)& 0.880 (1.2)& 0.816 (1.1) \\
			
                &\multirow{2}{*}{1200}& proposed& 1.000 (0.1) & 0.992 (0.1)& 0.984 (0.1)& 0.998 (0.5)& 0.980 (0.5)& 0.964 (0.4)& 0.990 (0.8)& 0.966 (0.7)& 0.936 (0.7) \\
				&&naive & 0.948 (0.2) & 0.888 (0.1)& 0.818 (0.1)& 0.920 (0.7)& 0.790 (0.6)& 0.726 (0.5)& 0.850 (1.0)& 0.700 (0.9)& 0.586 (0.8)\\
                \hline
				\multirow{6}{*}{$q = 0.25$}
				&\multirow{2}{*}{400}& proposed& 0.998 (0.3) & 0.992 (0.3)& 0.988 (0.2)& 0.996 (1.2)& 0.988 (1.1)& 0.980 (1.0)& 1.000 (1.9)& 0.992 (1.6)& 0.978 (1.5) \\
				&&naive  & 0.994 (0.3) & 0.980 (0.3)& 0.966 (0.3)& 0.996 (1.4)& 0.974 (1.2)& 0.970 (1.1)& 0.980 (2.1)& 0.950 (1.8)& 0.926 (1.7)\\
				&\multirow{2}{*}{800}& proposed& 0.998 (0.2) & 0.992 (0.2)& 0.986 (0.2)& 1.000 (0.9)& 0.988 (0.7)& 0.974 (0.7)& 0.996 (1.3)& 0.976 (1.1)& 0.964 (1.1) \\
				&&naive& 0.990 (0.2) & 0.966 (0.2)& 0.942 (0.2)& 0.972 (1.0)& 0.930 (0.8)& 0.900 (0.8)& 0.960 (1.5)& 0.904 (1.3)& 0.834 (1.2)  \\
			
                &\multirow{2}{*}{1200}& proposed& 1.000 (0.2) & 0.990 (0.1)& 0.980 (0.1)& 0.994 (0.6)& 0.976 (0.5)& 0.966 (0.5)& 0.986 (1.0)& 0.946 (0.9)& 0.920 (0.8)\\
				&&naive& 0.960 (0.2) & 0.900 (0.1)& 0.854 (0.1)& 0.912 (0.7)& 0.800 (0.6)& 0.728 (0.6)& 0.862 (1.1)& 0.712 (0.9)& 0.626 (0.9)  \\
                \hline
			\end{tabular}} }  }
	\end{table}
	
	\begin{table}[t]
		\caption{Empirical rejection probability(mean width of the uniform confidence band) for testing $H_0: g' = 0$ calculated from 500 Monte-Carlo simulations using Method~1 (Undersmoothing)}
		\label{table:RejProb}{\normalsize {\footnotesize \ \centering
		\resizebox{\textwidth}{!}{  
				\begin{tabular}{ccc|ccc|ccc|ccc}
					\hline
					&&&\multicolumn{3}{c}{DGP0}&\multicolumn{3}{c}{DGP1L}&\multicolumn{3}{c}{DGP1NL}\\
					&$N$  & Method &99\%& 95\%&90\%&99\%& 95\%&90\%&99\%& 95\%&90\%\\
					\hline
					\multirow{6}{*}{Average}
					&\multirow{2}{*}{400} &  proposed& 0.002 (1.2) & 0.050 (1.0)& 0.110 (0.9)& 0.088 (6.7)& 0.228 (5.6)& 0.396 (5.0)& 0.172 (7.1)& 0.420 (5.9)& 0.608 (5.3) \\
					&&naive& 0.038 (1.2) & 0.160 (1.0)& 0.288 (0.9)& 0.018 (6.5)& 0.062 (5.4)& 0.116 (4.9)& 0.032 (6.6)& 0.108 (5.5)& 0.194 (5.0)  \\
					&\multirow{2}{*}{800}&   proposed& 0.010 (0.8) & 0.032 (0.7)& 0.086 (0.6)& 0.154 (4.5)& 0.398 (3.8)& 0.538 (3.5)& 0.358 (4.6)& 0.676 (3.9)& 0.808 (3.6)
                    \\
					&&naive& 0.066 (0.8) & 0.212 (0.7)& 0.384 (0.6)& 0.012 (4.6)& 0.054 (3.9)& 0.114 (3.6)& 0.040 (4.7)& 0.124 (4.0)& 0.208 (3.6) \\	
                    &\multirow{2}{*}{1200} &  proposed& 0.006 (0.4) & 0.046 (0.3)& 0.114 (0.3)& 0.930 (1.2)& 0.956 (1.0)& 0.962 (0.9)& 0.888 (2.2)& 0.924 (1.9)& 0.956 (1.7)\\
					&&naive& 0.696 (0.4) & 0.850 (0.3)& 0.910 (0.3)& 0.524 (1.2)& 0.734 (1.0)& 0.814 (0.9)& 0.198 (2.1)& 0.412 (1.8)& 0.544 (1.6)  \\
                    \hline
					\multirow{6}{*}{$q= 0.45$}
					&\multirow{2}{*}{400}& proposed& 0.002 (2.1) & 0.016 (1.6)& 0.068 (1.5)& 0.090 (7.4)& 0.288 (5.9)& 0.442 (5.3)& 0.046 (11.6)& 0.220 (9.2)& 0.366 (8.3)  \\
					&&naive& 0.010 (2.0) & 0.068 (1.6)& 0.124 (1.4)& 0.022 (7.0)& 0.080 (5.5)& 0.150 (4.9)& 0.016 (10.0)& 0.046 (8.0)& 0.090 (7.2)\\
					&\multirow{2}{*}{800}& proposed& 0.002 (1.2) & 0.030 (1.0)& 0.056 (0.9)& 0.156 (4.9)& 0.440 (4.0)& 0.620 (3.6)& 0.122 (7.5)& 0.376 (6.2)& 0.510 (5.6) \\
					&&naive & 0.024 (1.2) & 0.104 (1.0)& 0.190 (0.9)& 0.022 (4.2)& 0.096 (3.5)& 0.168 (3.1)& 0.016 (6.9)& 0.054 (5.8)& 0.122 (5.2)\\
                    &\multirow{2}{*}{1200}& proposed& 0.010 (0.5) & 0.062 (0.4)& 0.112 (0.4)& 0.848 (1.7)& 0.900 (1.4)& 0.918 (1.3)& 0.768 (3.2)& 0.862 (2.6)& 0.894 (2.4) \\
					&&naive  & 0.400 (0.6) & 0.632 (0.5)& 0.750 (0.4)& 0.286 (1.9)& 0.514 (1.6)& 0.654 (1.4)& 0.090 (3.2)& 0.216 (2.6)& 0.336 (2.4)\\
					\hline
					\multirow{6}{*}{$q= 0.35$}
					&\multirow{2}{*}{400}& proposed& 0.002 (2.0) & 0.022 (1.6)& 0.042 (1.4)& 0.064 (8.5)& 0.172 (6.7)& 0.330 (6.0)& 0.050 (12.1)& 0.178 (9.6)& 0.282 (8.6)    \\
					&&naive & 0.020 (1.8) & 0.070 (1.4)& 0.106 (1.3)& 0.018 (7.2)& 0.058 (5.8)& 0.118 (5.2)& 0.006 (11.0)& 0.036 (8.9)& 0.084 (7.9)\\
					&\multirow{2}{*}{800}& proposed& 0.010 (1.3) & 0.028 (1.1)& 0.064 (1.0)& 0.130 (4.7)& 0.388 (3.9)& 0.572 (3.5)& 0.124 (7.5)& 0.340 (6.2)& 0.506 (5.6) \\
					&&naive& 0.032 (1.1) & 0.112 (0.9)& 0.188 (0.8)& 0.014 (4.7)& 0.102 (3.9)& 0.172 (3.5)& 0.016 (7.6)& 0.064 (6.3)& 0.112 (5.7)\\
                    &\multirow{2}{*}{1200}& proposed& 0.012 (0.5) & 0.062 (0.4)& 0.098 (0.4)& 0.834 (1.8)& 0.900 (1.5)& 0.936 (1.4)& 0.746 (3.5)& 0.862 (2.9)& 0.900 (2.6)\\
					&&naive & 0.380 (0.5) & 0.644 (0.4)& 0.758 (0.3)& 0.234 (1.9)& 0.456 (1.6)& 0.600 (1.4)& 0.080 (3.3)& 0.226 (2.7)& 0.332 (2.5\\
					\hline
					\multirow{6}{*}{$q = 0.25$}
					&\multirow{2}{*}{400}& proposed& 0.002 (2.2) & 0.018 (1.7)& 0.044 (1.5)& 0.036 (8.7)& 0.142 (6.9)& 0.268 (6.1)& 0.042 (12.5)& 0.138 (10.1)& 0.244 (9.0) \\
					&&naive& 0.010 (1.9) & 0.054 (1.5)& 0.112 (1.4)& 0.004 (7.8)& 0.050 (6.3)& 0.098 (5.6)& 0.000 (12.1)& 0.052 (9.6)& 0.084 (8.6)\\
					&\multirow{2}{*}{800}& proposed& 0.004 (1.2) & 0.034 (1.0)& 0.082 (0.9)& 0.104 (5.2)& 0.306 (4.3)& 0.482 (3.8)& 0.076 (8.4)& 0.280 (6.9)& 0.436 (6.2) \\
					&&naive& 0.024 (1.2) & 0.112 (1.0)& 0.176 (0.9)& 0.018 (5.1)& 0.078 (4.2)& 0.138 (3.8)& 0.008 (7.9)& 0.044 (6.6)& 0.106 (5.9)\\
                    &\multirow{2}{*}{1200}& proposed& 0.010 (0.6) & 0.044 (0.5)& 0.088 (0.4)& 0.822 (2.4)& 0.878 (2.0)& 0.928 (1.8)& 0.746 (3.7)& 0.850 (3.0)& 0.896 (2.7) \\
					&&naive & 0.276 (0.5) & 0.568 (0.4)& 0.696 (0.3)& 0.192 (1.9)& 0.432 (1.5)& 0.582 (1.4)& 0.080 (3.3)& 0.192 (2.7)& 0.302 (2.5)\\
					\hline
				\end{tabular}} }  }
		\end{table}

\begin{table}[t]
		\caption{Empirical rejection probability(mean width of the uniform confidence band) for testing $H_0: g' = 0$ calculated from 500 Monte-Carlo simulations uisng the Method~2 (Lepski)}
		\label{table:RejProb_Lepski}{\normalsize {\footnotesize \ \centering
		\resizebox{\textwidth}{!}{  
				\begin{tabular}{ccc|ccc|ccc|ccc}
					\hline
					&&&\multicolumn{3}{c}{DGP0}&\multicolumn{3}{c}{DGP1L}&\multicolumn{3}{c}{DGP1NL}\\
					&$N$  & Method &99\%& 95\%&90\%&99\%& 95\%&90\%&99\%& 95\%&90\%\\
					\hline
					\multirow{6}{*}{Average}
					&\multirow{2}{*}{400} &  proposed& 0.000 (0.5) & 0.000 (0.5)& 0.006 (0.4)& 0.622 (2.8)& 0.712 (2.5)& 0.746 (2.3)& 0.692 (3.1)& 0.766 (2.7)& 0.814 (2.5)  \\
					&&naive& 0.048 (0.6) & 0.112 (0.5)& 0.202 (0.5)& 0.000 (3.8)& 0.006 (3.3)& 0.012 (3.0)& 0.004 (3.8)& 0.024 (3.3)& 0.046 (3.0)\\
					&\multirow{2}{*}{800}&   proposed& 0.000 (0.4) & 0.002 (0.4)& 0.004 (0.4)& 0.698 (2.3)& 0.748 (2.0)& 0.778 (2.0)& 0.756 (2.6)& 0.806 (2.3)& 0.836 (2.1) 
                    \\
					&&naive& 0.160 (0.5) & 0.296 (0.4)& 0.378 (0.4)& 0.000 (2.9)& 0.006 (2.5)& 0.010 (2.4)& 0.010 (3.1)& 0.032 (2.7)& 0.058 (2.6) \\	
                    &\multirow{2}{*}{1200} &  proposed& 0.000 (0.3) & 0.004 (0.3)& 0.016 (0.3)& 0.826 (1.1)& 0.872 (1.0)& 0.906 (0.9)& 0.804 (2.0)& 0.852 (1.8)& 0.880 (1.7)\\
					&&naive& 0.470 (0.4) & 0.626 (0.3)& 0.706 (0.3)& 0.238 (1.4)& 0.428 (1.2)& 0.524 (1.1)& 0.032 (2.2)& 0.086 (2.0)& 0.158 (1.8) \\
                    \hline
					\multirow{6}{*}{$q= 0.45$}
					&\multirow{2}{*}{400}& proposed& 0.002 (0.6) & 0.002 (0.5)& 0.006 (0.4)& 0.620 (2.0)& 0.712 (1.7)& 0.772 (1.6)& 0.508 (3.2)& 0.644 (2.8)& 0.734 (2.5)  \\
					&&naive& 0.012 (0.6) & 0.058 (0.5)& 0.128 (0.5)& 0.014 (2.3)& 0.052 (2.0)& 0.078 (1.8)& 0.002 (3.7)& 0.016 (3.2)& 0.026 (3.0)\\
					&\multirow{2}{*}{800}& proposed& 0.000 (0.5) & 0.000 (0.4)& 0.002 (0.4)& 0.766 (1.7)& 0.838 (1.5)& 0.894 (1.3)& 0.692 (2.7)& 0.776 (2.4)& 0.836 (2.2) \\
					&&naive& 0.060 (0.5) & 0.138 (0.4)& 0.218 (0.4)& 0.024 (1.8)& 0.080 (1.6)& 0.150 (1.4)& 0.004 (3.3)& 0.010 (2.9)& 0.034 (2.7)\\
                    &\multirow{2}{*}{1200}& proposed& 0.000 (0.4) & 0.002 (0.3)& 0.006 (0.3)& 0.858 (1.2)& 0.894 (1.0)& 0.910 (0.9)& 0.818 (1.9)& 0.872 (1.6)& 0.896 (1.5) \\
					&&naive& 0.252 (0.4) & 0.472 (0.3)& 0.570 (0.3)& 0.124 (1.4)& 0.306 (1.2)& 0.414 (1.1)& 0.018 (2.1)& 0.062 (1.9)& 0.086 (1.7) \\
					\hline
					\multirow{6}{*}{$q= 0.35$}
					&\multirow{2}{*}{400}& proposed& 0.000 (0.6) & 0.002 (0.5)& 0.004 (0.5)& 0.458 (2.5)& 0.618 (2.1)& 0.688 (1.9)& 0.296 (3.7)& 0.516 (3.1)& 0.608 (2.9)   \\
					&&naive  & 0.010 (0.6) & 0.060 (0.5)& 0.108 (0.5)& 0.008 (2.5)& 0.032 (2.1)& 0.052 (2.0)& 0.000 (4.0)& 0.012 (3.4)& 0.022 (3.2)\\
					&\multirow{2}{*}{800}& proposed& 0.000 (0.6) & 0.000 (0.5)& 0.004 (0.4)& 0.692 (1.8)& 0.782 (1.6)& 0.820 (1.4)& 0.632 (2.9)& 0.738 (2.5)& 0.794 (2.3) \\
					&&naive& 0.052 (0.5) & 0.146 (0.5)& 0.232 (0.4)& 0.016 (2.0)& 0.076 (1.7)& 0.120 (1.6)& 0.002 (3.2)& 0.016 (2.8)& 0.032 (2.6)\\
                    &\multirow{2}{*}{1200}& proposed& 0.000 (0.3) & 0.002 (0.3)& 0.006 (0.3)& 0.852 (1.2)& 0.898 (1.0)& 0.912 (1.0)& 0.820 (2.1)& 0.856 (1.8)& 0.888 (1.7)    \\
					&&naive& 0.194 (0.3) & 0.396 (0.3)& 0.516 (0.3)& 0.096 (1.3)& 0.246 (1.1)& 0.342 (1.1)& 0.022 (2.4)& 0.060 (2.1)& 0.086 (2.0) \\
					\hline
					\multirow{6}{*}{$q = 0.25$}
					&\multirow{2}{*}{400}& proposed& 0.002 (0.6) & 0.006 (0.5)& 0.006 (0.5)& 0.308 (2.5)& 0.508 (2.1)& 0.610 (1.9)& 0.192 (3.8)& 0.388 (3.2)& 0.494 (3.0)\\
					&&naive& 0.004 (0.6) & 0.040 (0.5)& 0.076 (0.5)& 0.000 (2.7)& 0.022 (2.3)& 0.046 (2.1)& 0.002 (4.3)& 0.010 (3.7)& 0.024 (3.4)\\
					&\multirow{2}{*}{800}& proposed& 0.000 (0.5) & 0.006 (0.4)& 0.008 (0.4)& 0.612 (1.9)& 0.740 (1.6)& 0.808 (1.5)& 0.512 (3.4)& 0.676 (2.9)& 0.752 (2.7) \\
					&&naive& 0.036 (0.5) & 0.140 (0.4)& 0.208 (0.4)& 0.014 (2.2)& 0.056 (1.9)& 0.088 (1.7)& 0.008 (3.4)& 0.024 (2.9)& 0.040 (2.7) \\
                    &\multirow{2}{*}{1200}& proposed& 0.000 (0.4) & 0.010 (0.3)& 0.012 (0.3)& 0.830 (1.5)& 0.890 (1.3)& 0.916 (1.2)& 0.778 (2.7)& 0.830 (2.3)& 0.868 (2.2) \\
					&&naive& 0.170 (0.3) & 0.330 (0.3)& 0.444 (0.3)& 0.066 (1.5)& 0.194 (1.3)& 0.312 (1.2)& 0.012 (2.5)& 0.050 (2.2)& 0.070 (2.0)\\
					\hline
				\end{tabular}} }  }
		\end{table}
	
%Additionally, we provide information on the average width of the uniform confidence band across the grid points of $(t,z)$ and the 200 simulations. Comparing the quantiles, as the quantile becomes more extreme, the mean lengths generally increase due to decreasing local sample sizes. Consequently, the advantage of our proposed method diminishes for the extreme quantiles.

\begin{table}[t]
\caption{1000 $\times \text{Average Squared Bias}$  (Average Variance) of the undersmoothed estimators (Method~1) from 500 Monte-Carlo simulations over the grid $\mathcal{V}$}\label{table:BiasVariance}{\normalsize {\footnotesize \ \centering
\resizebox{.9\textwidth}{!}{  
\begin{tabular}{ccc|cc|cc|cc}
					\hline
					\multicolumn{3}{c}{} & \multicolumn{2}{c}{$\mathbf{DGP0}$}&  \multicolumn{2}{c}{$\mathbf{DGP1L}$}&  \multicolumn{2}{c}{$\mathbf{DGP1NL}$}\\
					&$N$  & Method& $\widehat{g}_{\widehat{h}_{\text{u}}}$ &  $\widehat{g^{\prime}}_{\widehat{h^{\prime}}_{\text{u}}}$ &  $\widehat{g}_{\widehat{h}_{\text{u}}}$ &  $\widehat{g^{\prime}}_{\widehat{h^{\prime}}_{\text{u}}}$ &  $\widehat{g}_{\widehat{h}_{\text{u}}}$ &  $\widehat{g^{\prime}}_{\widehat{h^{\prime}}_{\text{u}}}$ \\
					\hline
					\multirow{6}{*}{Average}
					&\multirow{2}{*}{400}&  proposed& 0.03 (3.95) & 0.14 (75.9)& 1.13 (119) & 5.65 (2509)& 1.26 (130) & 8.90 (2628)\\
					&&naive& 1.23 (4.19) & 7.73 (62.1)& 47.2 (120) & 286 (2242)& 62.9 (125) & 416 (2373)\\
					&\multirow{2}{*}{800}& proposed & 0.02 (2.05) & 0.07 (26.8)& 0.42 (61.6) & 3.33 (790)& 0.57 (64.4) & 2.72 (783)\\
					&&naive & 1.23 (2.14) & 7.33 (27.6)& 44.7 (63.8) & 273 (858)& 62.0 (66.1) & 407 (807)\\
                    &\multirow{2}{*}{1200}&  proposed& 0.00 (0.84) & 0.05 (36.0)& 0.04 (9.77) & 0.54 (303)& 0.12 (28.6) & 10.7 (590)\\
					&&naive& 1.19 (0.82) & 7.44 (53.8)& 19.4 (9.82) & 120 (141)& 60.9 (27.4) & 416 (671)\\
					\hline
					\multirow{4}{*}{$q = 0.45$}
					&\multirow{2}{*}{400}&  proposed & 0.19 (5.90) & 0.45 (140)& 1.84 (81.5) & 3.83 (1938)& 5.65 (195) & 10.9 (4662)\\
					&&naive & 1.37 (6.59) & 7.37 (117)& 22.0 (78.1) & 128 (1715)& 69.8 (175) & 427 (3293)\\
					&\multirow{2}{*}{800}& proposed& 0.08 (2.93) & 0.13 (50.9)& 0.68 (42.7) & 4.61 (939)& 1.77 (102) & 6.01 (2071)\\
					&&naive& 1.18 (2.84) & 6.83 (56.1)& 20.1 (39.7) & 116 (493)& 64.4 (96.5) & 408 (1616)\\
                    &\multirow{2}{*}{1200}&  proposed & 0.06 (1.20) & 0.06 (28.0)& 0.77 (15.8) & 0.52 (236)& 2.04 (43.1) & 8.80 (626)\\
					&&naive & 1.12 (1.22) & 6.42 (27.1)& 20.3 (16.5) & 118 (347)& 64.0 (41.0) & 423 (729)\\
					\hline
					\multirow{6}{*}{$q = 0.35$}
					&\multirow{2}{*}{400}&  proposed & 0.28 (6.37) & 0.38 (122)& 3.09 (112) & 6.38 (2208)& 8.12 (211) & 8.40 (4012)\\
					&&naive & 1.25 (5.54) & 6.22 (88.8)& 22.5 (91.3) & 129 (1654)& 70.0 (200) & 415 (3633)\\
					&\multirow{2}{*}{800}& proposed & 0.14 (2.93) & 0.31 (55.0)& 1.75 (45.4) & 4.10 (804)& 4.94 (107) & 9.91 (1981)\\
					&&naive& 1.11 (2.80) & 6.18 (33.7)& 21.1 (45.2) & 120 (701)& 68.8 (110) & 418 (1990)\\
                    &\multirow{2}{*}{1200}&  proposed& 0.09 (1.11) & 0.07 (28.3)& 1.13 (17.0) & 1.02 (296)& 3.74 (47.8) & 7.11 (1235)\\
					&&naive & 1.01 (1.12) & 5.80 (16.5)& 20.2 (17.9) & 120 (282)& 65.7 (46.2) & 419 (697)\\
					\hline
					\multirow{6}{*}{$q = 0.25$}
					&\multirow{2}{*}{400}&  proposed& 0.46 (6.88) & 0.44 (133)& 5.31 (120) & 4.40 (2111)& 14.1 (245) & 12.7 (4657)\\
					&&naive & 1.23 (5.85) & 6.07 (114)& 23.5 (99.4) & 125 (1711)& 71.7 (247) & 451 (4235)\\
					&\multirow{2}{*}{800}& proposed & 0.21 (2.85) & 0.23 (40.8)& 3.24 (51.3) & 6.68 (769)& 8.21 (122) & 12.5 (2858)\\
					&&naive & 1.05 (2.79) & 5.54 (38.6)& 22.6 (51.8) & 127 (678)& 70.8 (119) & 410 (1906)\\
                    &\multirow{2}{*}{1200}&  proposed& 0.14 (1.23) & 0.07 (27.9)& 2.22 (21.1) & 2.14 (628)& 5.61 (53.1) & 2.93 (793)\\
					&&naive& 0.93 (1.14) & 4.97 (17.9)& 20.8 (18.5) & 117 (257)& 66.7 (49.5) & 412 (711)\\
					\hline
\end{tabular}} }  }
\end{table}

\begin{table}[t]
\caption{1000 $\times \text{Average Squared Bias}$  (Average Variance) of the Lepski estimators (Method~2) from 500 Monte-Carlo simulations over the grid $\mathcal{V}$}\label{table:BiasVariance_Lepski}{\normalsize {\footnotesize \ \centering
\resizebox{.9\textwidth}{!}{  
\begin{tabular}{ccc|cc|cc|cc}
					\hline
					\multicolumn{3}{c}{} & \multicolumn{2}{c}{$\mathbf{DGP0}$}&  \multicolumn{2}{c}{$\mathbf{DGP1L}$}&  \multicolumn{2}{c}{$\mathbf{DGP1NL}$}\\
					&$N$  & Method& $\widehat{g}_{\widehat{h}}$ &  $\widehat{g^{\prime}}_{\widehat{h^{\prime}}}$ &  $\widehat{g}_{\widehat{h}}$ &  $\widehat{g^{\prime}}_{\widehat{h^{\prime}}}$ &  $\widehat{g}_{\widehat{h}}$ &  $\widehat{g^{\prime}}_{\widehat{h^{\prime}}}$ \\
					\hline
					\multirow{6}{*}{Average}
					&\multirow{2}{*}{400}&  proposed& 0.01 (0.72) & 0.01 (11.6)& 0.06 (19.1) & 0.62 (391)& 0.17 (19.3) & 18.6 (540)\\
					&&naive& 1.24 (0.86) & 7.56 (15.3)& 45.7 (23.8) & 279 (854)& 60.5 (24.7) & 422 (549)\\
					&\multirow{2}{*}{800}& proposed& 0.01 (0.39) & 0.02 (10.9)& 0.07 (10.5) & 0.94 (253)& 0.12 (11.6) & 11.2 (295)\\
					&&naive & 1.21 (0.42) & 7.48 (8.58)& 44.8 (12.5) & 276 (333)& 60.4 (13.1) & 414 (336)\\
                    &\multirow{2}{*}{1200}&  proposed& 0.00 (0.21) & 0.02 (8.47)& 0.01 (2.74) & 0.19 (90.4)& 0.17 (7.15) & 20.1 (293)\\
					&&naive& 1.18 (0.28) & 7.42 (5.81)& 19.3 (3.43) & 120 (84.5)& 59.8 (9.23) & 420 (237)\\
					\hline
					\multirow{4}{*}{$q = 0.45$}
					&\multirow{2}{*}{400}&  proposed & 0.10 (0.79) & 0.03 (7.65)& 2.19 (9.73) & 0.37 (100)& 7.39 (23.0) & 18.6 (273)\\
					&&naive & 1.25 (0.88) & 6.86 (7.42)& 21.8 (11.9) & 119 (123)& 69.5 (27.4) & 425 (317)\\
					&\multirow{2}{*}{800}& proposed& 0.05 (0.37) & 0.03 (6.64)& 0.69 (5.26) & 0.46 (82.3)& 2.68 (12.8) & 14.1 (240)\\
					&&naive& 1.14 (0.40) & 6.76 (5.41)& 19.5 (5.95) & 118 (86.2)& 63.0 (14.2) & 422 (364)\\
                    &\multirow{2}{*}{1200}&  proposed & 0.06 (0.25) & 0.02 (5.17)& 0.88 (3.56) & 0.28 (59.5)& 3.54 (7.96) & 26.9 (155)\\
					&&naive& 1.12 (0.29) & 6.47 (4.98)& 20.3 (4.09) & 118 (77.2)& 64.2 (9.20) & 435 (207)\\
					\hline
					\multirow{6}{*}{$q = 0.35$}
					&\multirow{2}{*}{400}&  proposed & 0.09 (0.77) & 0.04 (6.87)& 1.58 (11.8) & 0.60 (128)& 6.36 (28.4) & 15.2 (286)\\
					&&naive & 1.06 (0.88) & 6.18 (7.46)& 21.4 (14.1) & 124 (141)& 67.7 (33.1) & 431 (330)\\
					&\multirow{2}{*}{800}& proposed & 0.08 (0.37) & 0.06 (8.38)& 1.21 (5.90) & 0.59 (71.6)& 4.49 (14.0) & 14.3 (216)\\
					&&naive& 1.05 (0.40) & 6.10 (7.49)& 20.4 (6.46) & 121 (81.8)& 66.3 (15.5) & 438 (240)\\
                    &\multirow{2}{*}{1200}&  proposed& 0.08 (0.26) & 0.02 (8.39)& 1.32 (4.04) & 0.16 (57.0)& 5.31 (9.68) & 24.3 (181)\\
					&&naive & 1.00 (0.27) & 5.70 (3.63)& 20.3 (4.60) & 119 (58.8)& 65.3 (10.8) & 433 (223)\\
					\hline
					\multirow{6}{*}{$q = 0.25$}
					&\multirow{2}{*}{400}&  proposed& 0.10 (0.89) & 0.04 (8.54)& 1.89 (14.5) & 0.95 (105)& 7.42 (32.4) & 10.5 (227)\\
					&&naive & 0.94 (0.93) & 5.49 (8.73)& 21.3 (16.2) & 121 (120)& 67.8 (35.8) & 430 (328)\\
					&\multirow{2}{*}{800}& proposed & 0.10 (0.43) & 0.04 (6.02)& 1.62 (7.06) & 0.57 (73.1)& 5.93 (17.8) & 11.7 (488)\\
					&&naive & 0.96 (0.44) & 5.48 (6.42)& 21.1 (8.18) & 124 (104)& 66.1 (18.6) & 428 (478)\\
                    &\multirow{2}{*}{1200}&  proposed& 0.11 (0.27) & 0.02 (8.39)& 1.93 (4.82) & 0.39 (94.3)& 7.12 (12.5) & 16.8 (355)\\
					&&naive & 0.93 (0.29) & 5.02 (3.79)& 20.5 (5.11) & 117 (76.3)& 66.0 (12.1) & 427 (195)\\
					\hline
\end{tabular}} }  }
\end{table}

To gain further insight into the uniform inference results, we present the average squared bias and average variance across the grid $\mathcal{V}$ and the 500 simulation samples for the undersmoothed estimators in Tables~\ref{table:BiasVariance} and \ref{table:BiasVariance_Lepski}. These simulation results indicate that our proposed estimator consistently exhibits significantly smaller bias and comparable variance to the naive estimator. Specifically, the bias of our estimator is negligible relative to its variance, unlike the naive estimator. Furthermore, as the quantile approaches the median value, we observe a reduction in variance, while with increasing sample size or quantile level, both bias and variance diminishes consistently for the proposed method.

\begin{figure}[t]
{\centering
\includegraphics[width=0.25\textwidth]{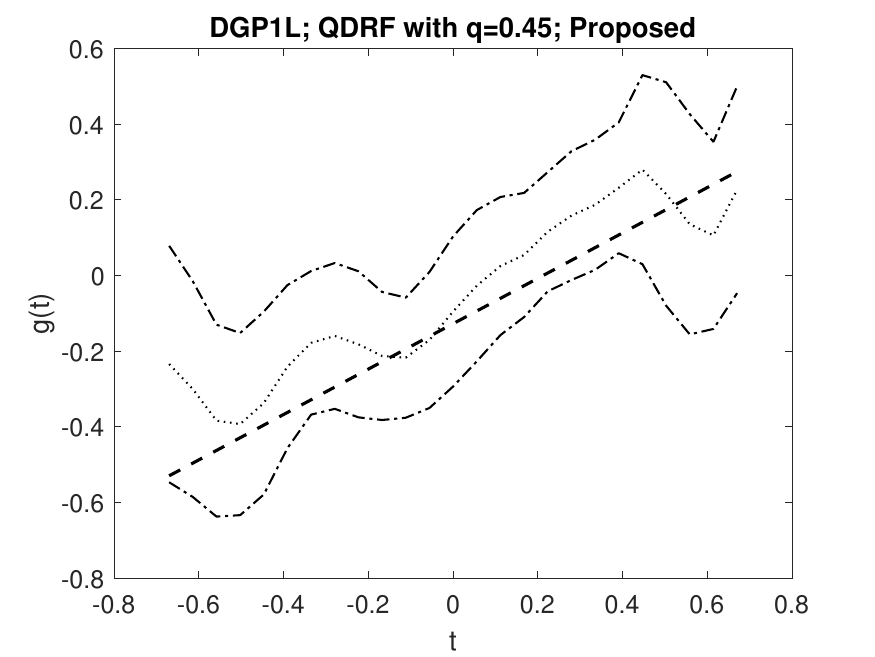}%
\includegraphics[width=0.25\textwidth]{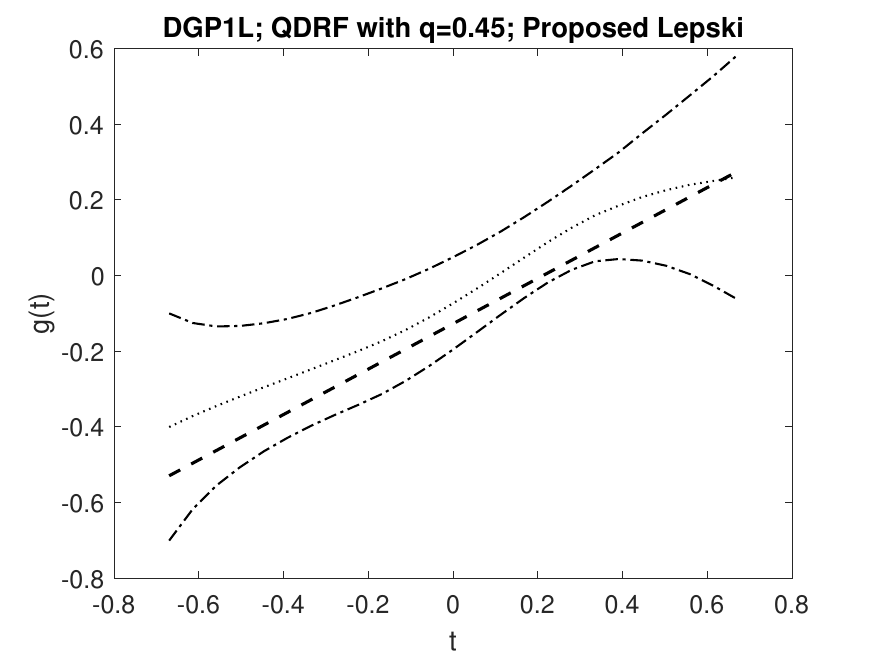}%
\includegraphics[width=0.25\textwidth]{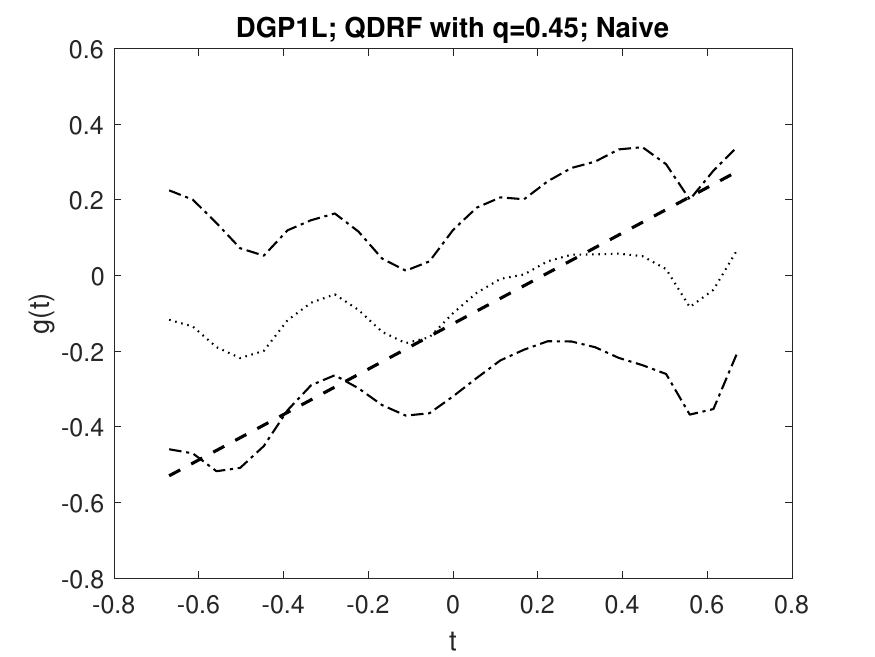}%
\includegraphics[width=0.25\textwidth]{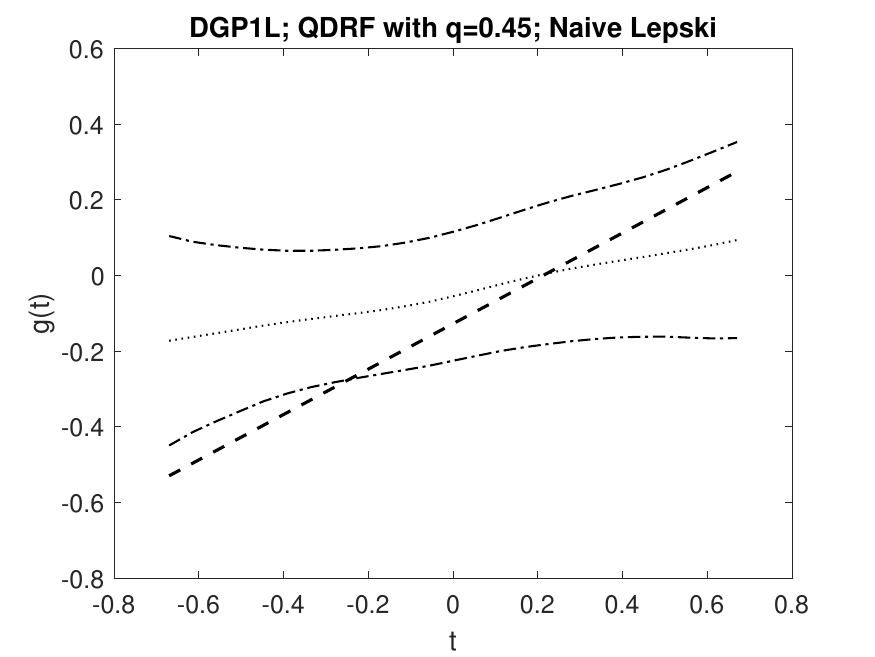}%

\includegraphics[width=0.25\textwidth]{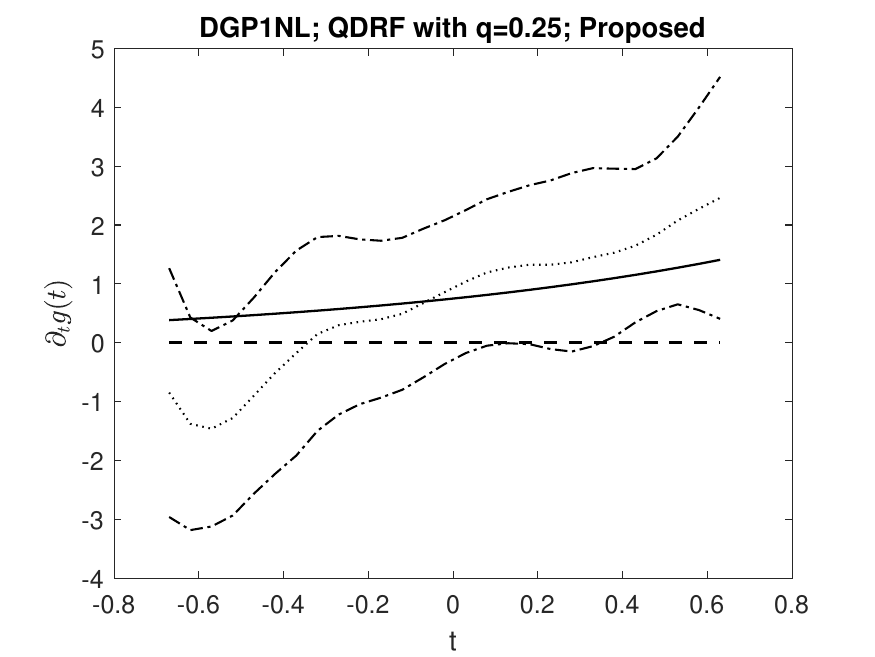}%
\includegraphics[width=0.25\textwidth]{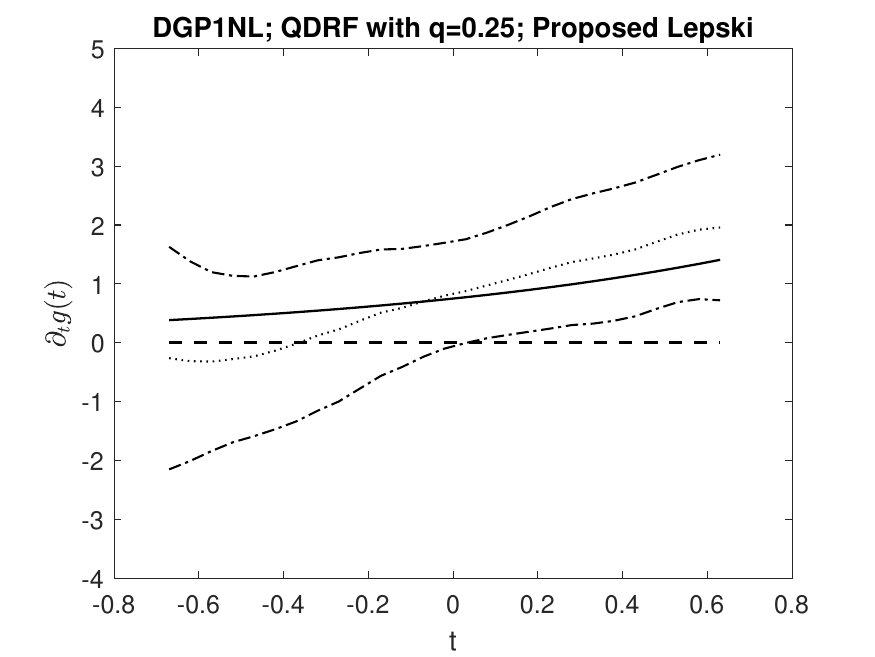}%
\includegraphics[width=0.25\textwidth]{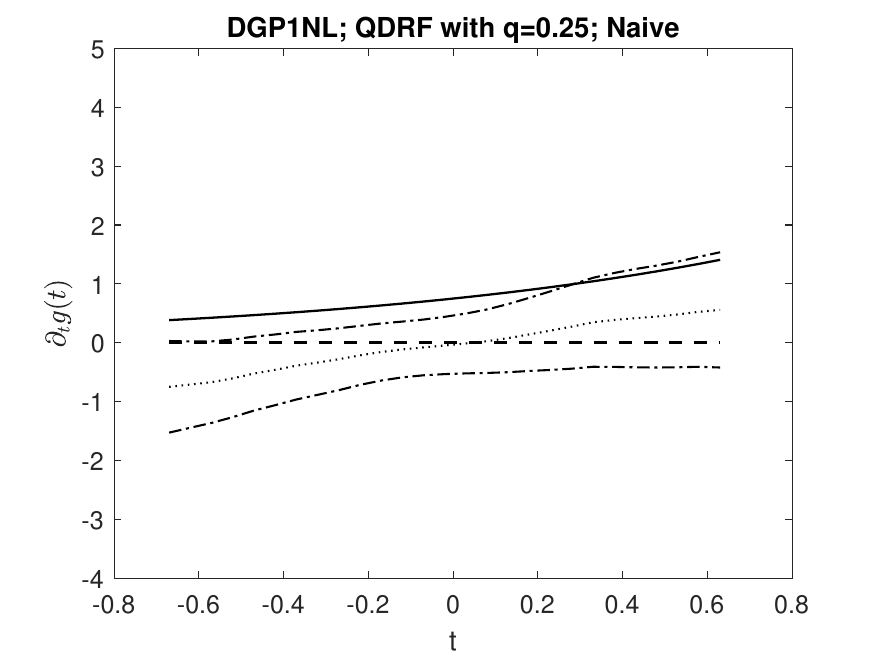}%
\includegraphics[width=0.25\textwidth]{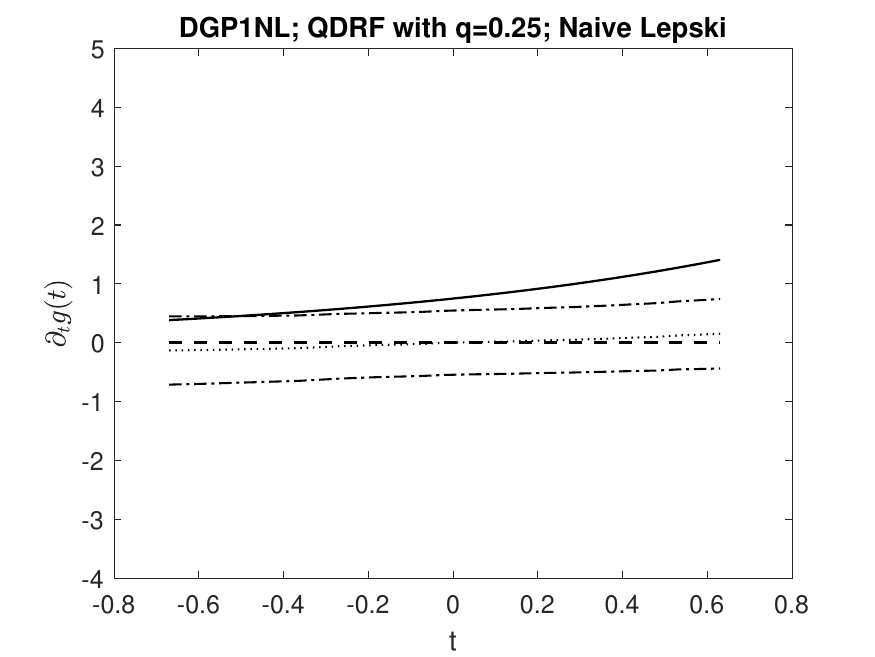}%
}
\caption{Plots of the estimators (dotted) of $g$ or $g^{'}$, with the 90\% uniform confidence band (dash-dotted) for $g$ and for testing $H_0:g^{'}=0$ (dashed), from samples with $N=1200$. The true $g^{'}$ curves are depicted by solid lines (Row~2).}
\label{fig:CB}
\end{figure}

To provide some visual intuition of the simulation results, Figures~\ref{fig:CB} depicts selected examples of the uniform 90\% confidence bands based on both the proposed and the naive methods for estimating $g$ from model $\mathbf{DGP1L}$ and $g'$ for $\mathbf{DGP1NL}$. These figures illustrate that the proposed method yields estimations of $g$ and $g'$ that align closely with the true curves, with the uniform confidence bands effectively covering the target. Conversely, the naive method introduces a noticeable bias due to confounding issues, resulting in confidence bands that miss the target.

\section{Empirical Study}

{\normalsize \label{sec:empirics} In this section, we apply the proposed
estimation method to analyze a dataset from the U.S. presidential campaign
to understand how the number of political advertisements aired causally
affects campaign contributions in non-competitive states. The dataset,
commonly utilized in continuous treatment effect literature, covers a range
of 0 to 22379 across $N=16265$ zip codes in non-competitive states 
\citep[see
e.g.][]{Urban_Niebler_2014, Fong_Hazlett_Imai_2018, ai2021unified,
huang2022unified}. }

{\normalsize The covariates $\boldsymbol{X}$ considered include $\log (\text{%
Population})$, the percentage of the population that is over 65 years old, $%
\log (\text{Median Family Income}+1)$, the percentage of the Hispanic
population, the percentage of the black population, $\log (\text{Population
density}+1)$, the percentage of college graduates and the indicator whether
the area can commute to a competitive state. Additional information can be
found in \cite{Fong_Hazlett_Imai_2018}. We normalize the covariates as $\tilde{\boldsymbol{X}}= \{\boldsymbol{X} - \min(\boldsymbol{X})\}/\{\max(\boldsymbol{X}) - \min(\boldsymbol{X})\}$.}

{\normalsize \cite{Urban_Niebler_2014} analyzed the causal relationship
between advertising and campaign contributions from this data using a binary
model. They compared the campaign contributions from the 5230 zip codes that
received more than 1000 advertisements with those from the remaining 11035
zip codes that received fewer than 1000 advertisements. Their analysis
revealed a significant causal effect of advertising in non-competitive
states on the level of campaign contributions. }

{\normalsize In contrast, \cite{ai2021unified} treated the number of
political advertisements as continuous and assumed a linear model on the
average response function. They applied log transformations to both the
outcome and treatment variables: $\log (\text{Contribution}+1)$ and $\log (\#%
\text{ads}+1)$, respectively, where \#ads is the number of advertisements.
Their analysis finds no significant causal effect of advertising on campaign
contributions, which is consistent with the finding in \cite%
{Fong_Hazlett_Imai_2018}. }

{\normalsize \cite{huang2022unified} proposed a unified framework for the
specification test of the continuous treatment effect models and rejected
the linear model assumed in \cite{ai2021unified}. They recommended a Tobit
model, combined with a Box-Cox transformation of the observed contributions
and a composite log transformation of \#ads, as a better fit for the data.
Specifically, their Box-Cox transformation is defined as $\text{BoxCox}(%
\text{Contribution},\lambda _{1},\lambda _{2})=\{(\text{Contribution}%
+\lambda _{2})^{\lambda _{1}}-1\}/\lambda _{1}$ with $(\tilde{\lambda}_{1},%
\tilde{\lambda}_{2})=(0.1397,0.0176)$. They then considered the observed
outcome variable, 
\begin{equation*}
Y^{\ast }(T)=Y=\text{BoxCox}(\text{Contribution},\tilde{\lambda}_{1},\tilde{%
\lambda}_{2})-\min \big\{\text{BoxCox}(\text{Contribution},\tilde{\lambda}%
_{1},\tilde{\lambda}_{2})\big\}\,,
\end{equation*}%
and the treatment variable, $T=\log (\log (\log (\#\text{ads}+1)+1)+2)$.
Their estimated Tobit model showed that campaign contributions increase
rapidly when $\#\text{ads}\in \lbrack 0,20]$, with improvements becoming
marginal thereafter. However, \cite{huang2022unified} did not give any
inferential results for the study. }

These conflicting findings from the previous studies indicate a complex causal relationship between advertising and campaign contributions, which parametric models may fail to capture. We use the same Box-Cox transformation for $Y$ as \cite{huang2022unified} but take $T = \{\tilde{T}-\min(\tilde{T})\}/\{\max(\tilde{T}) - \min(\tilde{T})\}$, where $\tilde{T}=\log(\#\text{ads} + 1 )$. Then, applying the proposed nonparametric method, we estimate the quantile dose-response function (QDRF), $g(t)=F^{-1}_{Y^*(t)}(q)$ and the partial derivative $\partial _{t}g$, along with 95\% uniform confidence bands, where $q=0.5$ and 0.75. 

\begin{figure}[tbp]
	\centering
   \includegraphics[width=0.25\textwidth]{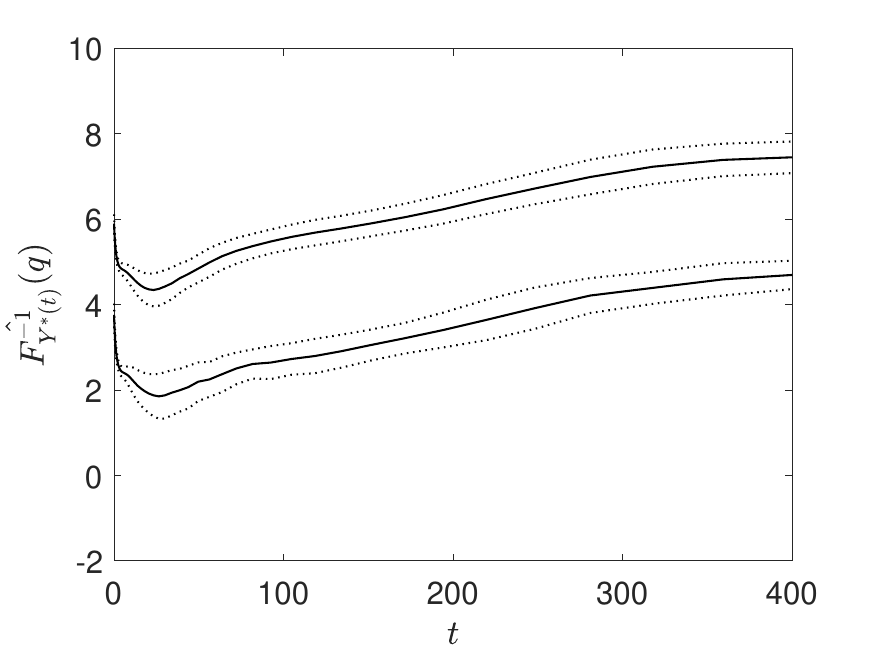}%
    \includegraphics[width=0.25\textwidth]{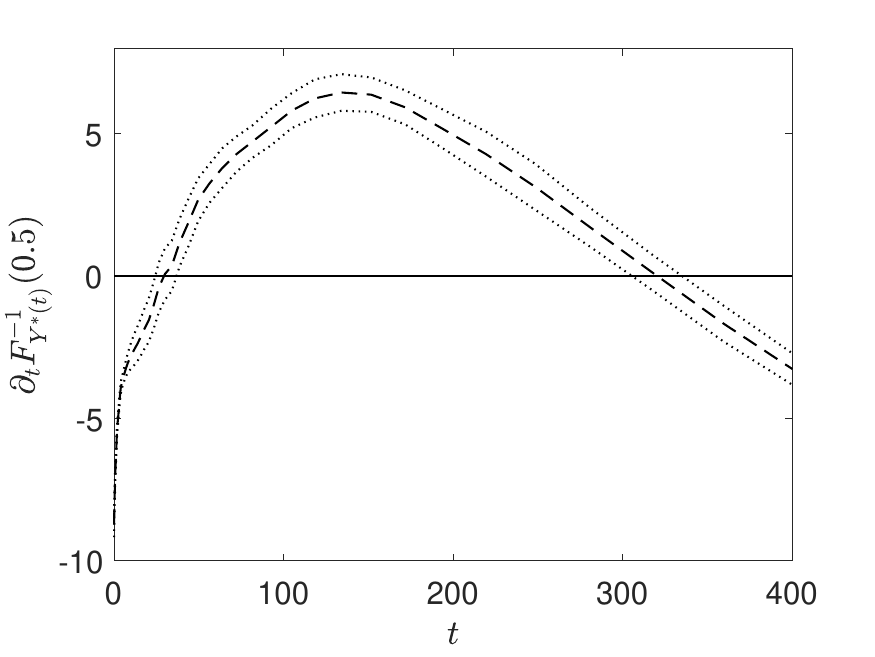}%
    \includegraphics[width=0.25\textwidth]{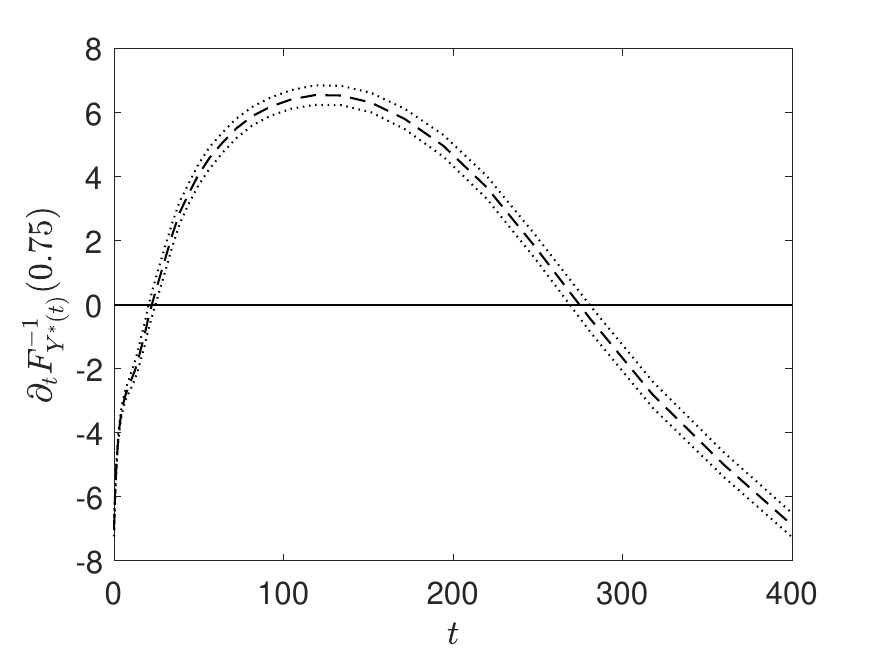}%
    \includegraphics[width=0.25\textwidth]{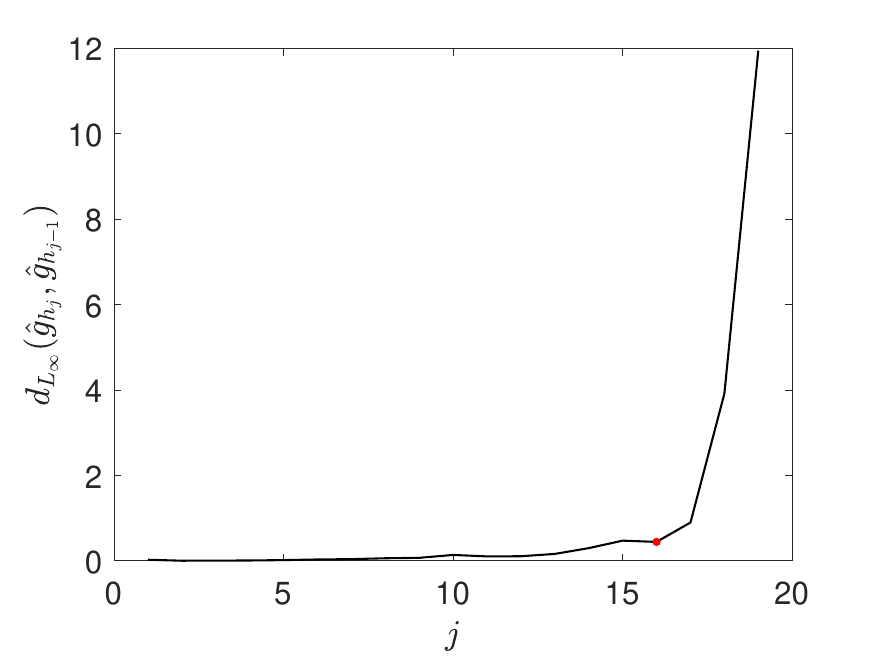}  
   
   \includegraphics[width=0.25\textwidth]{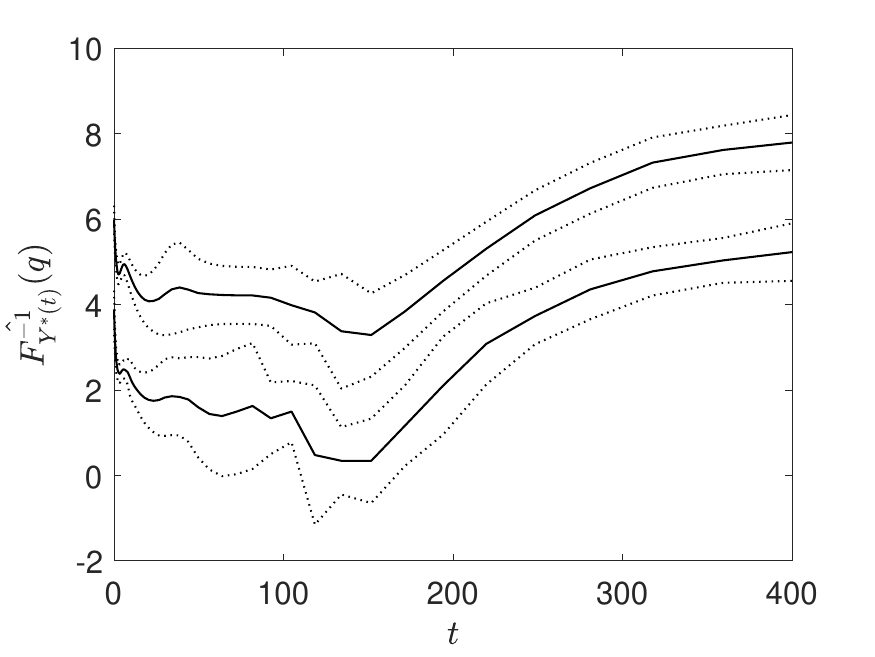}%
    \includegraphics[width=0.25\textwidth]{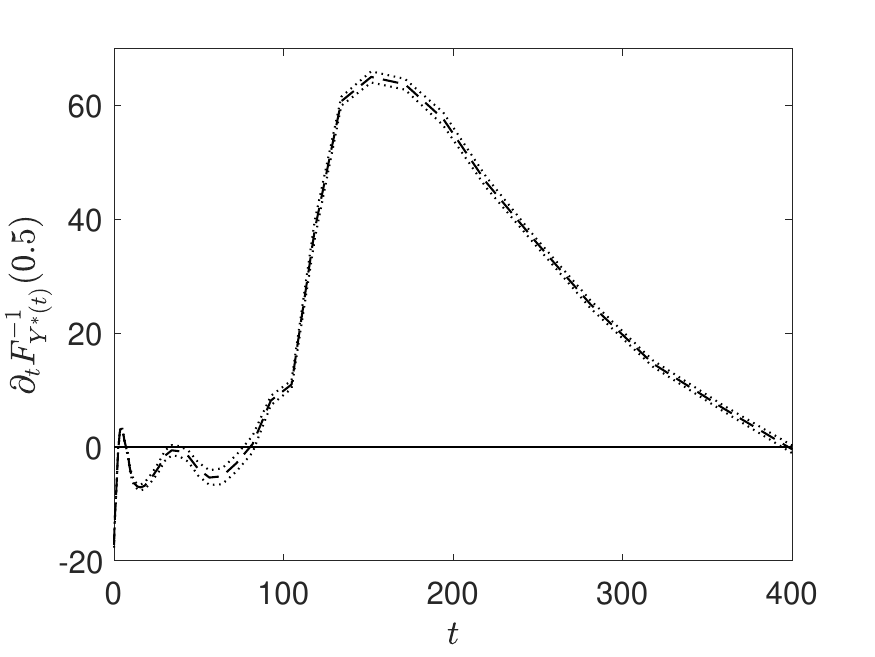}%
    \includegraphics[width=0.25\textwidth]{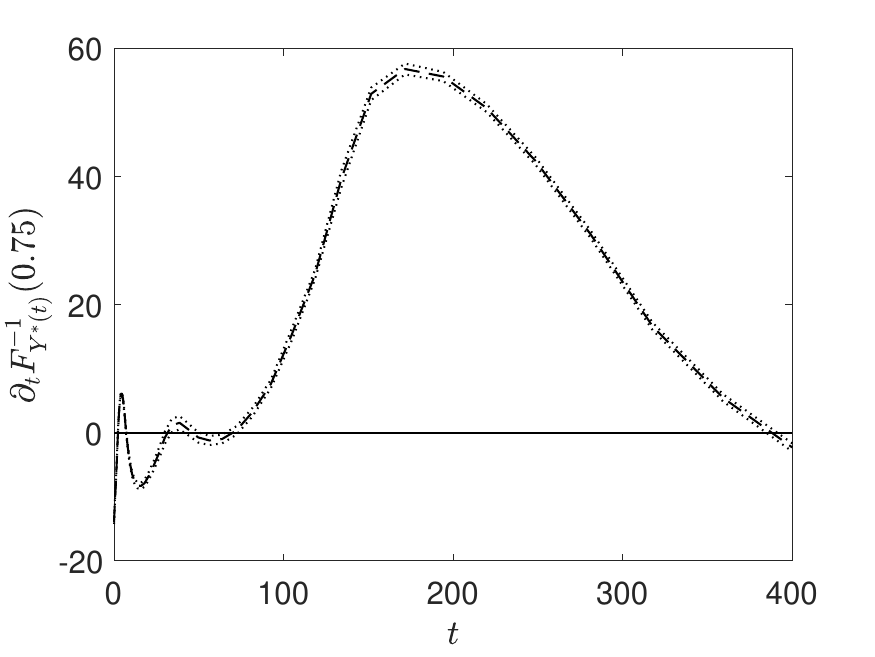}%
    \includegraphics[width=0.25\textwidth]{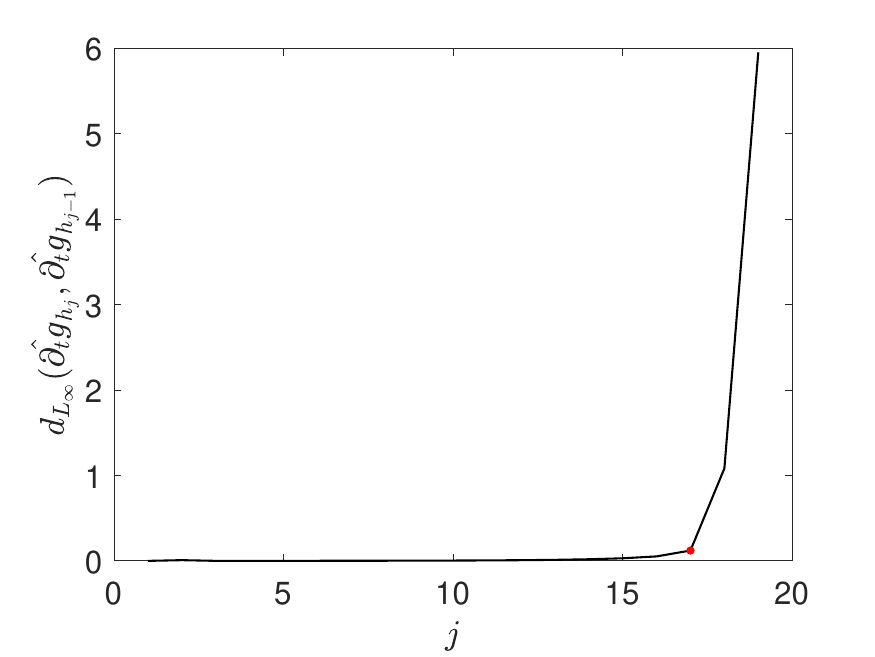}%
   \caption{Plots of the estimated QDRF with 95\% uniform confidence band ($q = 0.5$ and 0.75 from bottom to top) using Methods~1 (top left) and 2 (bottom left), the estimated first derivative of the QDRF with 95\% confidence band (2nd and 3rd column) using Methods~1 (top) and 2 (bottom), and the plots of $d_{L_\infty}(\widehat{g}_{h_j},\widehat{g}_{h_{j-1}})$ (top right) and $d_{L_\infty}(\widehat{\partial_t g}_{h_j},\widehat{\partial_t g}_{h_{j-1}})$ (bottom right) for Method~1, where $g$ is the median dose-response function. }
   \label{fig:USCampaign}
\end{figure}

Regarding the tuning parameters, we take $K_1=2$ and $K_2=1$ as \cite{ai2021unified} and \cite{huang2022unified}. The bandwidths are selected by the method described in Section~\ref{sec:tuning}. Specifically, the pilot bandwidth $h_0 = 1.1 \widetilde{h}$ with $\widetilde{h}$ selected by minimizing $G(2, 1, h)$ defined in Section~\ref{sec:tuning} w.r.t. $h$. The plots on the right of Figure~\ref{fig:USCampaign} depict the supremum distances over $j$ for the estimated median QDRF and its derivative, respectively. We choose the first $j$ where a significant increasing turning point appears in the plot, which gives $j=16$ for median dose-response $g$ and $j=17$ for the derivative. 

Figure~\ref{fig:USCampaign} presents the estimated QDRFs and their first derivatives, along with 95\% uniform confidence bands using Methods~1 and 2 introduced in Section~\ref{sec:NumericalDetails}. These derivatives and their confidence bands indicate no heterogeneity in the causal impact of advertising on the distribution of campaign contributions. Consistent with the ADRF results in \cite{ai2021unified} and \cite{huang2022unified}, the causal impact of advertising on the distribution of campaign contributions is significant for relatively small advertising volumes ($\#\text{ads} <400$) but becomes marginal beyond this threshold.

%\section*{Acknowledgments}
%The first author, Chunrong Ai acknowledges the financial support from the National Natural Science Foundation of China [grant number 71873138]. The last author, Zheng Zhang, acknowledges financial support from National Key R\&D Program of China (No. 2022YFA1008300),  the National Natural Science Foundation of China [grant number 12371284], and the Natural Science Foundation of Beijing [grant number 1222007]. We are grateful to Xiaohong Chen, Yingjie Feng, Haoze Hou, Hongjun Li, Jia Li, Liangjun Su, and seminar participants at National University of Singapore, Singapore Management University, and Tsinghua University for their comments and help on improving this manuscript. 

\clearpage
\appendix

\section*{Appendix}

\section{Proof of \eqref{eq:g0_min}}\label{app:iden_weighted}
 By Assumption \ref{as:TYindep}, i.e. $T\perp Y^*(t)|\bs{X}$, and using the tower property of conditional expectation, for every  $t\in\mathcal{T}$, the GDRF $g(t)$ can be point-wisely identified as follows: 
\begin{align*}
g(t)=&\arg \min_{a\in\mathbb{R}}\mathbb{E}\left[ \mathcal{L}\left( Y^{\ast }(t)-a\right)\right]\\
=&\arg \min_{a\in\mathbb{R}}\int\mathbb{E}\left[ \mathcal{L}\left( Y^{\ast }(t)-a\right)|\bs{X}=\bs{x}\right]f_{X}(\bs{x})d\bs{x}\\
=&\arg \min_{a\in\mathbb{R}}\int\mathbb{E}\left[ \mathcal{L}\left( Y^{\ast }(t)-a\right)|\bs{X}=\bs{x},T=t\right]f_{X}(\bs{x})d\bs{x} \quad (Assumption~\ref{as:TYindep})\\
=&\arg \min_{a\in\mathbb{R}}\int\mathbb{E}\left[ \mathcal{L}\left( Y-a\right)|\bs{X}=\bs{x},T=t\right]f_{X}(\bs{x})d\bs{x}\\
=&\arg \min_{a\in\mathbb{R}} \int\mathbb{E}\left[\pi(T,\bs{X}) \mathcal{L}\left( Y-a\right)|\bs{X}=\bs{x},T=t\right]\cdot f_{X|T}(\bs{x}|t)d\bs{x}\\
=&\arg \min_{a\in\mathbb{R}} \mathbb{E}\left[\pi(T,\bs{X}) \mathcal{L}\left( Y-a\right)|T=t\right].
\end{align*}

\section{Derivation of \eqref{def:pihat}}\label{app:pihat}
Let
$$
b_K:=\frac{1}{N(N-1)} \sum_{i=1}^N \sum_{j=1,i\neq j}^N u_K\left(T_i,\boldsymbol{X}_i\right). \quad
$$
Moreover, let $ U_{K \times N}:=\left(u_K\left(T_1,\boldsymbol{X}_1\right), \ldots, u_K\left(T_N,\boldsymbol{X}_N\right)\right) \in \mathbb{R}^{K \times N}$, $\boldsymbol{\pi}=(\pi_1,\ldots,\pi_{N})$,  and $F\left(\boldsymbol{\pi}\right):=$ $\sum_{i=1}^{N}(\pi_i-1)^2/2 $. Then  \eqref{E:cm2} can be written  as
\begin{equation}
	\left\{ 
	\begin{array}{cc}
		& \min _{\boldsymbol{\pi}} F\left(\boldsymbol{\pi}\right) \\[2mm] 
		& \text{subject to}\ U_{K \times N} \cdot \boldsymbol{\pi}=N\cdot b_K\,.%
	\end{array}
	\right.   \label{E:cm3}
\end{equation}
The conjugate convex function \citep{tseng1991relaxation} of $F(\cdot)$ is given by
$$
\begin{aligned}
	F^*(\boldsymbol{z}) & =\sup _{\boldsymbol{\pi}} \sum_{i=1}^N\left\{ z_i \pi_i- (\pi_i-1)^2/2 \right\}  =\sum_{i=1}^N\left\{z_i \pi_i^*-  (\pi^*_i-1)^2/2 \right\},
\end{aligned}
$$
where the $\pi_i^*$ 's satisfy the first order conditions:
$$
z_i=\pi_i^*-1 \Rightarrow \pi_i^*=z_i+1, \quad i=1, \ldots, N.
$$
Substuting $\pi_i^* = z_i+1$ into $F^*(\boldsymbol{z})$, we have
$$
F^*(\boldsymbol{z})=\sum_{i=1}^N\left\{  z_i(z_i+1)-z_i^2/2\right\}=\sum_{i=1}^N z_i^2/2+z_i=\sum_{i=1}^N \frac{1}{2}(z_i+1)^2-\frac{1}{2}.
$$
By \cite{tseng1991relaxation}, the dual problem of \eqref{E:cm3} is to take $\bs{z} = (-\gamma^*)^\top U_{K\times N}$ with $\gamma^*$ solving 
\begin{align*}
	&\max _{\gamma \in \mathbb{R}^K}\left\{-F^*\left(-\gamma^{\top} U_{K \times N}\right)+(-\gamma)^{\top} (N\cdot b_K)\right\} \\
	& =\max _{\gamma \in \mathbb{R}^K}\left[\sum_{i=1}^N \left\{-\frac{1}{2}(\gamma^\top u_K(T_i,\bs{X}_i)-1)^2 + \frac{1}{2}\right\}-\gamma^{\top} (N\cdot b_K)\right] \\
	& =\max _{\gamma \in \mathbb{R}^K} N\{\widehat{G}(\gamma) + 1/2\} ,
\end{align*}
where 
\begin{align*}
	\widehat{G}(\gamma)  = -\frac{1}{2N}\sum_{i=1}^N \left\{\gamma^{\top} u_K\left(T_i,\boldsymbol{X}_i\right)-1\right\}^2-\frac{1}{N(N-1)} \sum_{i=1}^N \sum_{j=1,i\neq j}^N \gamma^{\top}u_K\left(T_i,\boldsymbol{X}_i\right).
\end{align*}
Hence,
$$\widehat{\pi}_K(T_i,\boldsymbol{X}_i)=-\widehat{\gamma}^{\top}u_K(T_i,\boldsymbol{X}_i)+1,$$
where {\small\begin{align*}
		\widehat{\gamma}=&\arg\max	\widehat{G}(\gamma)\\
		=&\left(\frac{1}{N}\sum_{i=1}^Nu_K\left(T_i,\boldsymbol{X}_i\right)u_K^{\top}\left(T_i,\boldsymbol{X}_i\right)\right)^{-1}\left(\frac{1}{N}\sum_{i=1}^Nu_K\left(T_i,\boldsymbol{X}_i\right)-\frac{1}{N(N-1)} \sum_{i=1}^N \sum_{j=1,i\neq j}^N u_K\left(T_i,\boldsymbol{X}_i\right)\right).
\end{align*}}

\section{Verification of Assumption \ref{ass:vctype}(i)}

 \label{app:VC} 
 When $\mathcal{L}(v)=v^2$, we verify the
 Vapnik–Chervonenkis (VC) type conditions for $\mathcal{F}_1$ and $\mathcal{F}_2$ imposed in
 Assumption \ref{ass:vctype}(i). The existence of $M$ is guaranteed by Assumption %
\ref{ass:ciid} combined with the continuity of $g(t)$ and $%
g'(t)$. For fixed $a$ and $b$, the subgraph of the
function $(y,t)\mapsto\mathcal{L}(y-a-bt)$ satisfies 
\begin{align*}
&\{(y,t,u) : u\le \mathcal{L}(y-a-bt)\}=\{(y,t,u) : u\le(y-a-bt)^2\} \\
=& \{(y,t,u):y -a -bt \ge\sqrt{u\vee 0}\}\cup \{(y,t,u):y -a -bt \le-\sqrt{%
u\vee 0}\} \in \mathcal{G}\cup \mathcal{G},
\end{align*}
where
\begin{align*}
\mathcal{G}:=\{\{(y,t,u):x_1\cdot y + x_2\cdot t + x_3\cdot \sqrt{u \vee 0} +
x_4\le 0\}:x_1,x_2,x_3,x_4\in\mathbb{R}\},
\end{align*}
is the set generated by the four measurable functions: $(y,t,u)\mapsto y$, $%
(y,t,u)\mapsto t$, $(y,t,u)\mapsto \sqrt{u \vee 0}$ and $(y,t,u)\mapsto 1$.
By \citet[Lemma 2.6.15, Lemma 2.6.18 (iii)]{van1996weak}, $\mathcal{G}$ is a
VC class of set. Then by \citet[Lemma 2.6.17 (iii)]{van1996weak}, $\mathcal{G%
}\cup\mathcal{G}$ is also a VC class of set. Hence, $\mathcal{F}_1$ in
Assumption\ref{ass:vctype}(i) is a VC-subgraph class. By 
\citet[Theorem
2.6.7]{van1996weak}, the VC-type condition for $\mathcal{F}_1$ imposed in
Assumption \ref{ass:vctype}(i) holds. Similarly, the VC-type condition for $%
\mathcal{F}_2$ imposed in Assumption \ref{ass:vctype}(i) holds. 

 When $\mathcal{L}(v) = v\{\tau-\mathds{1}(v\leq 0)\}$, the VC-type
conditions for $\mathcal{F}_1$ and $\mathcal{F}_2$ can be similarly
verified.

\bibliographystyle{chicago}
\bibliography{Semiparametric}
 
\end{document}